\documentclass[aps,pre,groupedaddress,showpacs,showkeys,floatfix,twocolumn]{revtex4}

\usepackage{graphicx}

\begin{document}
\title{Generalized models as a universal approach to the analysis of nonlinear dynamical systems}
\author{Thilo~Gross}
\email[]{thilo.gross@physics.org}
\affiliation{AG Nichtlineare Dynamik, Universit\"at Potsdam, am Neuen Palais 10, 14469 Potsdam, Germany}
\author{Ulrike~Feudel}
\affiliation{ICBM, Carl von Ossietzky Univerit\"at, PF 2503, 26111 Oldenburg, Germany}
\date{\today}

\begin{abstract}
We present a universal approach to the investigation of the dynamics in generalized models. 
In these models the processes that are taken into account are not restricted to specific functional forms.
Therefore a single generalized models can describe a class of systems which share a similar structure.
Despite this generality, the proposed approach allows us to study the dynamical properties of generalized models 
efficiently in the framework of local bifurcation theory. The approach is based on a normalization procedure that is
used to identify natural parameters of the system. The Jacobian in a steady state is then derived as a function 
of these parameters. The analytical computation of local bifurcations using computer algebra reveals conditions 
for the local asymptotic stability of steady states and provides certain insights on the global dynamics of the system. 
The proposed approach yields a close connection between modelling and nonlinear dynamics. We illustrate 
the investigation of generalized models by considering examples from three different disciplines of science: 
a socio-economic model of dynastic cycles in china, a model for a coupled laser system and a general ecological food web.
\end{abstract}

\maketitle

\section{Introduction \label{secIntro}}
Dynamical systems are used to study phenomena from diverse disciplines of science 
such as laser physics, population ecology, socio-economic studies and many more.
The corresponding mathematical models have often the form of balance equations, 
in which the time evolution of the state variables is determined by gain and loss terms. 
Depending on the processes that are taken into account, a single state variable can 
be effected by several gains and losses. In the modeling process the modeler usually
decides first which processes are important and need to be included in the model. These
processes determine the structure of the model. In the second step each of the terms 
is described by a specific mathematical function, which can be based on theoretical 
reasoning or empirical evidence. In this way a \emph{specific model} for the phenomenon 
under consideration is constructed. The investigation of specific models is a powerful approach,
that has, in many cases, revealed interesting insights. However, the specific mathematical
functions on which these models are based depend critically on the modelers knowledge of 
the system. While the modeler may have much information about certain processes, others may 
be known or believed to exist which are very difficult to quantify.  

Specific models are therefore often based on a large number of assumptions. This can be illustrated very well by considering 
an example from ecology: The so-called Holling type-II function is regularly used to describe the 
predator-prey interaction in food webs and food chains\cite{Holling:Characteristics}. This function takes major biological 
effects into account. But, it can not possibly capture all the subtle biological details that exist in nature. 
Such details may involve the formation of swarms to confuse predators, adaptation of the prey 
to high predator densities, density dependent changes in the predator's foraging strategy to name a few. 
In models such details are often omitted on purpose since the modeler is interested 
in general insights that do not depend on specific properties of the system under consideration. 
But on the other hand, certain details may have a strong impact on the qualitative behavior of the system. 
For instance, it has been shown that even minor variations in the functional form of the 
predator-prey interaction can have a strong impact on the system's stability \cite{EnrichmentPaper,Bernd:Sensitivity}.         

In contrast to specific functional forms, the basic structure of the system is much 
easier to determine. For instance in our ecological example it is easier to say that the predation depends 
on the population density of predator and prey than to derive the exact functional form that quantifies 
this dependence. It can therefore be useful to consider \emph{generalized models} which describe 
the structure of a system in terms of gains and losses but do not restrict these processes to specific functional forms.
If a system is considered in which some processes are known with a high degree of certainty 
while others remain uncertain it can be useful to study partially generalized models in which 
some processes are described by specific functional forms, while others remain general. In other 
cases fully generalized models in which all processes are modeled by generalized functions can be advantageous. 
Generalized models describe systems with a higher generality than specific models. They depend on 
less assumptions and enable the researcher to consider the system from an abstract point of view.

Despite the generality of generalized models, it is possible to compute the stability of a steady state 
in these models analytically. This reveals the exact relationship between the qualitative features 
of functional forms and the local bifurcations of steady states \cite{EnrichmentPaper}. 
In this way the application of generalized models enables us to investigate the local stability of 
systems with a high degree of generality. Moreover, we can draw certain 
conclusions on the global dynamics of the system under consideration. For instance, the presence 
of chaotic or quasiperiodic dynamics as well as the existence of homoclinic bifurcations can be deduced 
form certain local bifurcations of higher codimension. In this way the presence of chaotic dynamics 
can be shown for a whole class of model systems \cite{ExponentPaper}. 

In this paper we present a universal approach to the investigation of generalized models based on balance equations. 
We illustrate this approach using three models from three different fields of science: socio-economics, laser physics 
and population dynamics. These examples demonstrate that the proposed approach can be used to study a wide range of 
systems. 
 
The paper is organised as follows. We introduce our approach by considering a simple socio-economic model
in Sec.~\ref{secSocial}. The underlying principles of the normalization procedure are discussed in more detail
while we study a coupled laser system in Sec.~\ref{secLasers}. Finally, we show that the proposed 
approach is applicable to more involved problems by investigating a model for general food webs 
in Sec.~\ref{secFoodWeb}. 

\section{A generalized model of dynastic cycles \label{secSocial}}
Chinese history is characterised by repeated transitions from despotism to anarchy and back \cite{Usher:DynasticCycles}. 
This periodic behavior is known as the dynastic cycle. As a young dynasty rises from anarchy 
a period of order and prosperity begins. As a result the population grows. At 
some point the aging dynasty can not control the growing population any longer. The results 
are poverty, crime and civil war which lead back to a period of anarchy. 
By contrast, European dynasties exhibit stable behavior for extended periods of time.
 
It has been shown in \cite{Usher:DynasticCycles} that the existence of dynastic cycles as well as 
steady state behavior (e.g. stable dynasties) can be predicted by a simple three-variable 
model that describes the evolution of a society of farmers $X$, bandits $Y$ and rulers $Z$.
The dynamics of this model have been investigated in \cite{Feichtinger:DynasticCycles}.  
In these investigations a specific model was used in which the interactions between 
the different social groups were described by specific functions. In the following 
we study a generalized version of this model in which the interactions are 
described by general functions. 

\subsection{Formulation of a generalized model of dynastic cycles}
In the model the growth of the population of farmers is described by a function 
$S(X)$. This function represents agricultural production, but also losses because of 
natural mortality including diseases. The population  
of farmers is reduced by crime $C(X,Y)$ and taxes $T(X,Z)$. Since the bandits profit 
from the crime their number is assumed to increase proportional to $C(X,Y)$. In addition 
crime increases the willingness of the farmers to support rulers. Therefore 
the number of rulers also grows proportional to $C(X,Y)$. The rulers fight crime 
by reducing the number of bandits. This effect of the rulers on the bandit population 
is described by a function $L(Y,Z)$. Finally, we take into account that 
the numbers of rulers and bandits decrease because of retirement and natural mortality.
This loss is denoted by $M(Y)$ for the bandits and $R(Z)$ for the rulers.
Taking all these factors into account one obtains differential equations of the form
\begin{equation}
\label{eqSocialModel}
\begin{array}{r c l}
\dot{X}&=&S(X)-C(X,Y)-T(X,Z),\\
\dot{Y}&=&\eta C(X,Y) - L(Y,Z) -M(Y),\\
\dot{Z}&=&\nu C(X,Y) - R(Z)
\end{array}
\end{equation}
where $\eta$ and $\nu$ are constant factors.

\subsection{Normalization of the socio-economic model}
So far we have formulated a generalized version of the model proposed in \cite{Usher:DynasticCycles}. 
In general the investigation of such a model would start with the computation of steady states.
However, in our model we can not compute steady states with the chosen degree of generality.
Instead, we assume that at least one steady state exists. This assumption is in general 
valid in a large parameter range. We denote the population densities in this steady state 
by $X^*$, $Y^*$ and $Z^*$ respectively. 

Let us now consider the local asymptotic stability of the steady state under consideration.  
In a system of autonomous ordinary differential equations (ODEs), such as Eq.~(\ref{eqSocialModel}), the stability 
of steady states depends on the eigenvalues of the corresponding Jacobian matrix \cite{Guckenheimer:DynamicalSystems}. 
For any system ODEs of the form $\dot{\rm \bf x}={\rm \bf g}({\rm \bf x})$ the Jacobian ${\rm \bf J}$ can be computed as 
\begin{equation}
\label{eqDefJ}
J_{m,n}=\left. \frac{\partial g_m}{\partial x_n} \right|_{\rm \bf x=x^*} \hspace{1cm}m,n=1,\ldots,N.
\end{equation} 
In principle we could therefore compute the Jacobian of our system directly by applying this definition 
to the system from Eq.~(\ref{eqSocialModel}).
However, the Jacobian obtained in this way would depend explicitly on the unknown steady state. 
Because of this dependence the interpretation of any insights gained from such a Jacobian is difficult. 
In order to avoid this difficulty we study a normalized system. By means of normalization it possible to 
arrive at a system for which the Jacobian does not depend explicitly on the unknown steady state. 
Instead, the unknown steady state only appears in certain parameters which are easily interpretable. 

In the following we use the abbreviated notation 
\begin{equation}
\label{eqSocialNotation}
\begin{array}{r c l r c l}
S^*&:=&S(X^*), 
C^*&:=&C(X^*,Y^*),\\
T^*&:=&T(X^*,Z^*), 
L^*&:=&L(Y^*,Z^*),\\
M^*&:=&M(Y^*), 
R^*&:=&R(Z^*).
\end{array}
\end{equation}
Using this notation, we define normalized state variables
\begin{equation}
\label{eqSocialVar}
\displaystyle x\displaystyle =\displaystyle\frac{\displaystyle X}{\displaystyle X^*}, 
\displaystyle y\displaystyle =\displaystyle\frac{\displaystyle Y}{\displaystyle Y^*}, 
\displaystyle z\displaystyle =\displaystyle\frac{\displaystyle Z}{\displaystyle Z^*}
\end{equation}
and normalized functions
\begin{equation}
\label{eqSocialFunc}
\begin{array}{r c l}
\displaystyle s(x)&:=&\displaystyle \frac{S(X^*x)}{S^*},\\[5pt]
\displaystyle c(x,y)&:=&\displaystyle \frac{C(X^*x,Y^*y)}{C^*},\\[5pt]
\displaystyle t(x,z)&:=&\displaystyle \frac{T(X^*x,Z^*z)}{T^*},\\[5pt]
\displaystyle l(y,z)&:=&\displaystyle \frac{L(Y^*y,Z^*z)}{L^*},\\[5pt]
\displaystyle m(y)&:=&\displaystyle \frac{M(Y^*y)}{M^*},\\[5pt]
\displaystyle r(z)&:=&\displaystyle \frac{R(Z^*z)}{R^*}.
\end{array}
\end{equation}
Note that the normalized quantities have been defined in such a way that
\begin{equation}
\label{eqSocialIdentity}
x^*=y^*=z^*=s^*=c^*=t^*=l^*=m^*=r^*=1
\end{equation}
holds.
 
Let us now rewrite our model in terms of the normalized variables. We start by substituting 
the definitions Eq.~(\ref{eqSocialVar}) and Eq.~(\ref{eqSocialFunc}) into Eq.~(\ref{eqSocialModel}). 
We obtain
\begin{equation} 
\label{eqSocialInterModel}
\begin{array}{r c l}
\dot{x}&=&\displaystyle \frac{S^*}{X^*}s(x)-\frac{C^*}{X^*}c(x,y)-\frac{T^*}{X^*}t(x,z),\\[5pt]
\dot{y}&=&\displaystyle \frac{\eta C^*}{Y^*}c(x,y)-\frac{L^*}{Y^*}l(y,z)-\frac{M^*}{Y^*}m(y),\\[5pt]
\dot{z}&=&\displaystyle \frac{\nu C^*}{Z^*}c(x,y)-\frac{R^*}{Z^*}r(z).
\end{array}
\end{equation}  

In order to simplify these equations we consider the system in the steady state. Because of Eq.~(\ref{eqSocialIdentity})
this yields
\begin{equation} 
\label{eqSocialInterModelII}
\begin{array}{r c l}
0&=&\displaystyle \frac{S^*}{X^*}-\frac{C^*}{X^*}-\frac{T^*}{X^*},\\[5pt]
0&=&\displaystyle \frac{\eta C^*}{Y^*}-\frac{L^*}{Y^*}-\frac{M^*}{Y^*},\\[5pt]
0&=&\displaystyle \frac{\nu C^*}{Z^*}-\frac{R^*}{Z^*}.
\end{array}
\end{equation} 
Since Eq.~(\ref{eqSocialInterModelII}) contains only constants the equation has to hold even if the system 
is in a non-stationary state. It is therefore reasonable to define 
\begin{equation}
\label{eqSocialDefAlpha}
\begin{array}{r c l}
\alpha_{\rm x}&=&\displaystyle \frac{S^*}{X^*} = \frac{C^*}{X^*}+\frac{T^*}{X^*},\\[5pt]
\alpha_{\rm y}&=&\displaystyle \frac{\eta C^*}{Y^*} = \frac{L^*}{Y^*}+\frac{M^*}{Y^*},\\[5pt]
\alpha_{\rm z}&=&\displaystyle \frac{\nu C^*}{Z^*} =\frac{R^*}{Z^*}.
\end{array}
\end{equation} 
Note, that the third line of Eq.~(\ref{eqSocialDefAlpha}) allows us to replace all constant 
factors in the third line of Eq.~(\ref{eqSocialInterModel}) by $\alpha_{\rm z}$. In order
to replace the other constant factors in the model in a similar way we define
\begin{equation}
\begin{array}{r c l}
\label{eqSocialDefBeta}
\beta_{\rm x}&:=&\displaystyle\frac{1}{\alpha_{\rm x}} \frac{C^*}{X^*},\\[5pt]
\beta_{\rm y}&:=&\displaystyle\frac{1}{\alpha_{\rm y}} \frac{L^*}{Y^*},\\[5pt]
\tilde{\beta_{\rm x}}&:=&\displaystyle 1-\beta_{\rm x}=\frac{1}{\alpha_{\rm x}} \frac{T^*}{X^*},\\[5pt]
\tilde{\beta_{\rm y}}&:=&\displaystyle 1-\beta_{\rm y}=\frac{1}{\alpha_{\rm y}} \frac{M^*}{Y^*}.
\end{array}
\end{equation}
In the following we call $\alpha_{\rm x}$, $\alpha_{\rm y}$, $\alpha_{\rm z}$, 
$\beta_{\rm x}$ and $\beta_{\rm y}$ \emph{scale parameters}.
The definitions Eq.~(\ref{eqSocialDefAlpha}) and Eq.~(\ref{eqSocialDefBeta}) effectively group 
the gain and loss terms together. As we will show below it is advantageous to define the 
scale parameters in this way since it results in easily interpretable parameters. 

By applying Eq.~(\ref{eqSocialDefAlpha}) and Eq.~\ref{eqSocialDefBeta} we can write 
Eq.~(\ref{eqSocialInterModel}) as 
\begin{equation}
\label{eqSocialNormalized}
\begin{array}{r c l}
\dot{x}&=&\alpha_{\rm x}(s(x)-\beta_{\rm x}c(x,y)-\tilde{\beta_{\rm x}}t(x,z)),\\
\dot{y}&=&\alpha_{\rm y}(c(x,y)-\beta_{\rm y}l(y,z)-\tilde{\beta_{\rm y}}m(y)),\\
\dot{z}&=&\alpha_{\rm z}(c(x,y)-r(z)).
\end{array}
\end{equation}
As a result of the normalization we have obtained a normalized model with the same structure as
the original model. But, we know that the steady state under consideration 
is located at $x^*=y^*=z^*=1$ in the normalized model. One could argue that the normalization 
procedure is merely a transformation of difficulties: In order to perform 
the normalization we had to introduce the scale parameters that again depend on the unknown steady state.
However, in the following we show that the scale parameters can be easily interpreted in the context of the model. 
That means, the values of the scale parameters can be determined by measurements or theoretical reasoning 
rather than explicit computation. In this way the normalization procedure provides us with a set of 
parameters, depending on which the dynamics of the system can be studied. 
Let us now consider the interpretation of these parameters in more detail.

From the way in which $\alpha_{\rm x}$, $\alpha_{\rm y}$ and $\alpha_{\rm z}$ appear
in Eq.~(\ref{eqSocialNormalized}) it can be guessed that these parameters
denote timescales. This guess is confirmed if one considers Eq.~(\ref{eqSocialDefAlpha}). 
In this equation $\alpha_{\rm x}$ is defined as the total production rate divided 
by the number of farmers. In other words this parameter denotes the total per-capita 
production rate of the farmers in the steady state. The parameter $\alpha_{\rm y}$ 
is defined as the per-capita growth rate of the bandits. At the same time it has to 
be the per-capita rate at which bandits disappear (to prison, retirement or death).
Consequently, we could write $\alpha_{\rm y}=1/\tau_{\rm y}$, where $\tau_{\rm y}$ 
is the typical length of a bandits career. In the same way $\alpha_{\rm z}$ is related 
to the typical length of a rulers career. 

The parameters $\beta_{\rm x}$ and $\beta_{\rm y}$ quantify the relative importance 
of certain processes. We have defined $\beta_{\rm x}$ as the per-capita loss rate 
of farmers because of crime divided by the total per-capita loss rate.
This means that $\beta_{\rm x}$ is simply the fraction of the farmer's losses that 
occurs because of crime. Likewise $\tilde{\beta_{\rm x}}$ is the fraction of the farmer's losses 
that occurs because of taxes. In a similar way, the parameter $\beta_{\rm y}$ denotes the fraction of the 
loss rate of bandits that occurs because of interaction with the rulers. In other words 
$\beta_{\rm y}$ denotes the fraction of bandits that are eventually caught. At the same time
$\tilde{\beta_{\rm y}}$ is the fraction of bandits that escape the rulers till they 
retire or die because of natural mortality.  

\subsection{The Jacobian of the socio-economic model}
So far we have derived a normalized model in which the location of the 
steady state under consideration is known. We have shown that the constant factors that appear 
in the normalization can be identified as meaningful parameters. Let us now compute the 
Jacobian of this normalized model. In the Jacobian all nonvanishing 
derivatives of the normalized functions with respect to the state variables appear.
We consider these derivatives as additional parameters. In the following 
these parameters are called \emph{exponent parameters}. The exponent parameters
are defined as 
\begin{equation}
\label{eqSocialDefExponents}
\begin{array}{r c l r c l}
s_{\rm x} &:=&\displaystyle \left. \frac{\partial}{\partial x} s(x)   \right|_{{\bf x}=1} 
c_{\rm x} &:=&\displaystyle \left. \frac{\partial}{\partial x} c(x,y) \right|_{{\bf x}=1}\\[5pt]
c_{\rm y} &:=&\displaystyle \left. \frac{\partial}{\partial y} c(x,y) \right|_{{\bf x}=1} 
t_{\rm x} &:=&\displaystyle \left. \frac{\partial}{\partial x} t(x,z) \right|_{{\bf x}=1}\\[5pt]
t_{\rm z} &:=&\displaystyle \left. \frac{\partial}{\partial z} t(x,z) \right|_{{\bf x}=1} 
l_{\rm y} &:=&\displaystyle \left. \frac{\partial}{\partial y} l(y,z) \right|_{{\bf x}=1}\\[5pt]
l_{\rm z} &:=&\displaystyle \left. \frac{\partial}{\partial z} l(y,z) \right|_{{\bf x}=1}
m_{\rm y} &:=&\displaystyle \left. \frac{\partial}{\partial y} m(y)   \right|_{{\bf x}=1}\\[5pt]
r_{\rm z} &:=&\displaystyle \left. \frac{\partial}{\partial z} r(z)   \right|_{{\bf x}=1}
\end{array}
\end{equation}
In order to understand the nature of the exponent parameters it is useful to consider 
the effect of the normalization on a specific function. For instance in the specific model 
studied by \cite{Usher:DynasticCycles,Feichtinger:DynasticCycles} the natural mortality rate of the bandits is modeled
by the function
\begin{equation}
M(Y)=AY,
\end{equation} 
where $A$ is a constant parameter. According to our normalization procedure (see Eq.~(\ref{eqSocialFunc})) the
corresponding normalized function is simply
\begin{equation}
m(y)=\frac{AY}{AY^*}=y.
\end{equation}
Therefore the exponent parameter is $m_{\rm y}=1$ regardless of the value of $A$.
In a similar way it can be shown that any specific function of the form 
\begin{equation}
M(Y)=AY^q
\end{equation}
corresponds to an exponent parameter
\begin{equation}
m_{\rm y}=q.
\end{equation}
Since we have normalized all functions in the same way, this does not only hold for the mortality 
term $M(Y)$ but for all functions in the model. For all processes that can be modeled with monomial 
expressions the exponent parameters are therefore identical to the exponent of the mononomial.
If a process is not described by a mononomial then the corresponding exponent parameter still measures
the nonlinearity of the process in the steady state. Consider for instance the specific function
\begin{equation}
C(X,Y)=\frac{AXY}{K+X}
\end{equation}
that is employed in \cite{Usher:DynasticCycles} to describe crime. In this case our normalization procedure 
yields the exponent parameter
\begin{equation}
c_{\rm x}=\frac{1}{1+X^*/K}. 
\end{equation}      
That means the parameter is almost one for $X^* \ll K$ where $C(X^*,Y^*)$ is almost linear in $X^*$. But, it 
vanishes for $X^* \gg K$ where $C(X^*,Y^*)$ is almost constant in $X^*$. For more examples 
of exponent parameters for specific functional forms see \cite{Thesis}.  

In this paper we will not assume that $C(X,Y)$ or $M(Y)$ have any specific functional form. 
But, we will use the insights gained from the consideration of specific funtional forms 
to interpret the exponent parameters. This interpretation reveals realistic ranges and values 
for these parameters. 

For instance, the parameter $s_{\rm x}$ measures the nonlinearity of the total production rate
as a function of the number of farmers. In a country in which empty land of sufficient quality is 
available we can assume that the productivity increases linearly
with the number of farmers. In this case we have $s_{\rm x}\approx 1$. However, if the production is not 
limited by the number of farmers, but by the lack of usable land then the production is a constant function 
of the number of farmers. This corresponds to $s_{\rm x}=0$. We can therefore interpret $s_{\rm x}$ as an 
indicator for the availability of usable land in the steady state.

Using the scale and exponent parameters the Jacobian of the system can be written as 
\begin{equation}
\begin{array}{c}
\label{eqSocialJac}
{\rm \bf J}=\left(\begin{array}{c c c} 
\alpha_{\rm x} & & \\ & \alpha_{\rm y} & \\ & & \alpha_{\rm z} 
\end{array}\right) \times\\
\left(\begin{array}{c c c}
s_{\rm x}-\beta_{\rm x}c_{\rm x}-\tilde{\beta_{\rm x}}t_{\rm x} & -\beta_{\rm x}c_{\rm y} & -\tilde{\beta_{\rm x}}t_{\rm z}\\
c_{\rm x} & c_{\rm y}-\beta_{\rm y}l_{\rm y}-\tilde{\beta_{\rm y}}m_{\rm y} & -\beta_{\rm y}l_{\rm z}\\
c_{\rm x} & c_{\rm y} & -r_{\rm z}  
\end{array}\right).
\end{array}
\end{equation}
One of the parameters $\alpha_{\rm x}, \alpha_{\rm y}, \alpha_{\rm z}$ can always be set to one by means of 
timescale normalization. This leaves us with 13 parameters. For comparison the specific model proposed in 
\cite{Usher:DynasticCycles} contains 11 parameters.

\subsection{Computation of bifurcations}
\begin{figure}
\centering
\includegraphics[width=8cm,height=6cm]{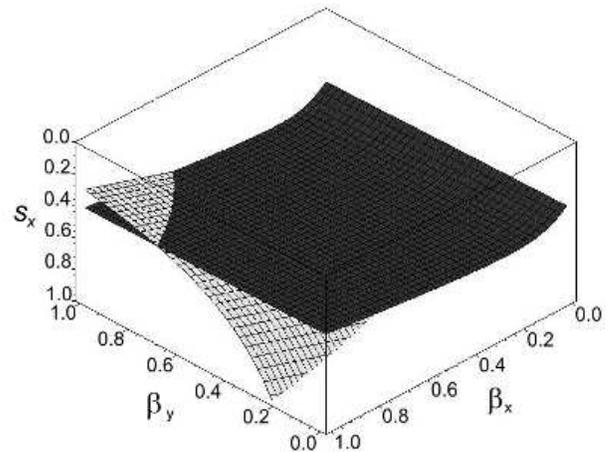}
\caption{Bifurcations in a society of farmers, bandits and rulers. Stable dynasties occur in the 
topmost volume of the parameter space shown. If the availability of usable 
land $s_{\rm x}$ is increased the stability is lost as a Hopf bifurcation (dark surface) or a bifurcation of saddle-node type 
(light surface) is crossed. The Hopf bifurcation gives rise to oscillatory behavior that is known as the dynastic cycle.
The fraction of farmer's losses that occur because of crime $\beta_{\rm x}$ and the fraction of the bandits 
that get eventually caught $\beta_{\rm y}$ have only a weak effect on the stability of the dynasty. \label{figSocial}} 
\end{figure}

The Jacobian which we have obtained in the previous section enables us to compute the stability of the steady 
states. In general the stability properties of dynamical systems change abruptly at critical parameter values
which are known bifurcation points\cite{Guckenheimer:DynamicalSystems,Kuznetsov:Elements}. 
Let us focus on the bifurcations that are encountered as one parameter is 
changed. Of these local codimension-1 bifurcations only two types exist.  The first type 
is characterised by the presence of a single zero eigenvalue of the Jacobian. In this paper we refer to these 
bifurcations collectively as \emph{bifurcations of saddle-node type}. These bifurcations generally correspond 
to the emergence or destruction of equilibiria or the exchange of stability properties between two equilibria. 
Some examples of this type of bifurction are the saddle-node 
bifurcation, the transcritical bifurcation or the pitchfork bifurcation. The bifurcations of saddle-node type 
can be found analytically by demanding that the determinant of the Jacobian vanishes. The other type of 
local codimension-1 bifurcation is the Hopf bifurcation. A Hopf bifurcation point is characterised by the presence 
of two purely imaginary eigenvalues of the Jacobian in the steady state. In this bifurcation the steady state 
becomes unstable as a limit cycle emerges or vanishes. In order to compute this bifurcation we use the 
method of resultants \cite{HopfPaper}. Although this method yields explicit 
analytical results the corresponding formulas are too lengthy to be presented here. 
Instead, the bifurcations are shown in a three-parameter bifurcation diagram in Fig.~\ref{figSocial}. 

In the figure the steady 
state under consideration is stable in the top-most volume of the parameter space. The stability of the steady 
state is lost in a Hopf bifurcation or in a bifurcation of saddle-node type. Since these bifurcations are of codimension one, 
the corresponding bifurcation points form hyper-surfaces in the parameter space. For the model under 
consideration the Hopf bifurcation surface is of particular interest. As this bifurcation surface is crossed 
the steady state becomes unstable while a limit cycle emerges. This gives rise to the dynastic cycle that is 
observed in simulations. The fact that the bifurcation surface lies almost perpendicular to the direction 
of $s_{\rm x}$ indicates that this parameter is of greater importance for the stability of the system than $\beta_{\rm x}$ or $\beta_{\rm y}$. 
Although such insights should be confirmed in a more careful mathematical investigation, they illustrate 
that bifurcation analysis of generalized models can be used to identify important parameters in a system. 

The scale and exponent parameters which we have defined above capture only the local behavior of the generalized model. 
The bifurcation analysis which we can perform is therefore limited to the investigation of local bifurcations.
Nevertheless, this analysis enables us to draw certain conclusions on the global dynamics of the 
generalized model. For example in Fig.~\ref{figSocial} there is a line in which the Hopf bifurcation 
surface intersects the bifurcation surface of saddle-node type. The points on such a line are codimension-2 
Gavrilov-Guckenheimer bifurcation points. The existence of these points implies that a region of 
complex (chaotic or quasiperiodic) dynamics exists generically in the model \cite{Kuznetsov:Elements}. 
In fact, it is shown in \cite{Feichtinger:DynasticCycles} that chaotic dynamics appear in a specific version 
of the generalized model considered here.  
  
In summary we can say that the investigation of generalized models allows us to generalize insights 
from specific models. In particular, generalized models can be used to investigate the local dynamics
(e.g. emergence of dynastic cycle from a Hopf bifurcation), to identify important parameters (e.g. 
availability of land $s_{\rm x}$) and to draw certain conclusions on the global dynamics (e.g. the existence 
of complex dynamics).

\section{A generalized model of coupled lasers \label{secLasers}}
In the previous section we have shown that the local bifurcations of generalized models 
can be computed efficiently if the model is normalized in a certain way. This approach can be 
summarized in form of the following 
algorithm:
\begin{enumerate}
  \item{Formulation of a generalized model}
  \item{Normalization}
    \begin{itemize} 
    \item{Define normalized variables and functions}
    \item{Rewrite model in terms of the normalized quantities}
    \item{Identify scale parameters}
    \end{itemize}
  \item{Computation of the Jacobian}
    \begin{itemize}
    \item{Define exponent parameters}
    \item{Interpret the exponent parameters} 
    \item{Compute the entries of the Jacobian matrix}
    \end{itemize}
  \item{Bifurcation analysis, \ldots}
\end{enumerate}
We argue that essentially the same procedure can be applied to a large variety of different systems which can be 
described in the form of balance equations.
In order to demonstrate this point we consider a system of coupled lasers. Although the scientific background 
is entirely different, it turns out that the laser model can be treated in the same way as the socio-economic model. 
For the analysis presented here the most striking difference between the physical and the socio-economic model 
is the amount of available information. In the socio-economic model all processes are difficult to quantify 
and are therefore candidates for generalization. By contrast, the underlying quantum mechanical foundations of 
laser physics are well understood. Many of the processes in the model can therefore be quantified 
with a high degree of certainty. Nevertheless, it can be useful to generalize certain processes 
which can be in principle quantified, but depend on many details of the specific system under consideration.
In the following we show that the algorithm outlined above can be applied in such a partially 
generalized system.

\subsection{Formulation of a model for coupled lasers}
Models from laser physics are among the classical applications of the theory 
of dynamical systems \cite{Haken:Lasers,Mandel:Lasers,Lauterborn:Lasers}. 
In particular coupled lasers are known to exhibit interesting dynamics.
The investigation presented here is inspired by the theoretical and experimental results 
reported in \cite{Fabiny:Lasers,Roy:Lasers}.
In this work it is shown that the dynamics of two coupled single-mode lasers can be described by 
equations of the form
\begin{equation}
\label{eqLaserModel}
\begin{array}{r c l}
\dot{A}_1&=&e_1 A_1 G_1 + r_1 G_1 + \kappa\sqrt{A_1 A_2}\cos(\delta)-L_1(A_1)\\
\dot{A}_2&=&e_2 A_2 G_2 + r_2 G_2 + \kappa\sqrt{A_1 A_2}\cos(\delta)-L_2(A_2)\\
\dot{G}_1&=&p_1- s_1 G_1 -r_1G_1 -e_1 A_1 G_1 \\
\dot{G}_2&=&p_2- s_2 G_2 -r_2G_2 -e_2 A_2 G_2 \\
\dot{\delta}&=& \frac{\kappa}{\tau_c} \left( \sqrt{\frac{A_1}{A_2}} + \sqrt{\frac{A_2}{A_1}} \right) \sin(\delta)  -\delta_0 \\
\end{array}
\end{equation} 
In order to discuss these equations, let us use the index $i=\{1,2\}$ to refer to laser 
1 and 2 respectively. In the equations the intensity of laser $i$ is described by the variable $A_i$ while 
the gain of each laser is described by the variable $G_i$. The variable $\delta$ denotes the phase difference 
between laser $1$ and laser $2$. 

The equations that governs the time evolution of $G_i$, the gain of laser $i$, contains four terms. 
The first term $p_i$ represents a constant pumping of the laser. The second term $s_iG_i$ describes the decrease
of the gain because of spontaneous emission of photons which do not go into the laser mode. The term 
$r_iG_i$ represents spontaneous emission into the laser mode. The final term $e_iA_iG_i$ corresponds to stimulated 
emission. The quantities $p_i$, $s_i$, $r_i$ and $e_i$ that appear in these equations are assumed to be constants.

Spontaneous or stimulated emission of photons into the laser mode of laser $i$ increases the intensity $A_i$.
This is described by the first two terms in the intensity equations. The third terms in these equations models 
the optical coupling of the lasers. The specific functional form of these terms as well as the equation that 
governs the evolution of the phase difference are results of Stratonovich calculus \cite{Fabiny:Lasers}. 
In these equations the constant $\kappa$ describes the strength of the coupling. The parameter $\tau_c$ is 
the cavity round trip time and $\delta_0$ is the difference of the phase velocity of the lasers without coupling. 

The only general function that appears in the model equations is $L_i(A_i)$. This function 
is used to describe the loss of laser intensity. Photons can be lost from the laser mode because of 
the absorbtion in the medium or because of escape from the resonator. Nonlinear optical effects 
like multi-photon absorbtion introduce nonlinearities in the intensity dependence of absorbtion 
inside the resonator \cite{Politi:Nonlinearities}. Other nonlinear effects, like for instance temperature dependence or 
intensity dependence of the refractive index can effect the goodness of the resonator. 
As a consequence the losses of intensity can depend nonlinearly on the intensity. 
While all of these effects can be quantified for a specific laser medium, this would 
yield a complicated model that describes only one very specific type of laser. In many models 
studied in literature it is instead assumed that the losses depend linearly on the intensity of the 
laser.  

\subsection{Normalization of the physical model}
Let us now normalize the generalized model formulated above. We start by assuming that there is at least one steady state.
The dynamical variables in the steady state are denoted by ${A_1}^*,{A_2}^*,{G_1}^*,{G_2}^*$ and $\delta^*$ respectively.
Furthermore, we denote the values of the generalized functions in the steady state by ${L_i}^*$. As in the previous example 
we define the normalized state variables 
\begin{equation}
\label{eqLaserVars}
\begin{array}{r c l}
\displaystyle a_i&:=& \displaystyle\frac{A_i}{{A_i}^*} \\[5pt]
\displaystyle g_i&:=& \displaystyle\frac{G_i}{{G_i}^*} \\
\end{array}
\end{equation}
and the normalized function
\begin{equation}
\label{eqLaserFunc}
l_i(a_i):=\displaystyle\frac{L_i({A_i}^*a_i)}{{L_i}^*} \\
\end{equation}

Normalization of the variable $\delta$ is not necessary. We could argue that $\delta$ is already 
an easily observable and interpretable variable, therefore the steady state value $\delta^*$
can be treated as a scale parameter. However in the following it will turn out that this reasoning is not necessary.

In the next step we rewrite the model in terms of the normalized 
variables. By substitution of Eq.~(\ref{eqLaserVars}) and Eq.~(\ref{eqLaserFunc})
into Eq.~(\ref{eqLaserModel}) we obtain 
\begin{equation}
\label{eqLaserLong}
\begin{array}{r c l}
\displaystyle \dot{a}_i&=& \displaystyle\frac{e_i {A_i}^*{G_i}^*}{{A_i}^*} a_i g_i + \frac{r_i {G_i}^*}{{A_i}^*} g_i 
              + \frac{\kappa\sqrt{{A_1}^*a_1{A_2}^*a_2}}{{A_i}^*}\cos(\delta)\\
              & &\displaystyle -\frac{{L_i}^*}{{A_i}^*}l_i(a_i)\\[5pt]
\displaystyle \dot{g}_i&=& \displaystyle\frac{p_i}{{G_i}^*}- \frac{s_i {G_i}^*}{{G_i}^*} g_i- \frac{r_i {G_i}^*}{{G_i}^*} g_i- 
             \frac{e_i {A_i}^* {G_i}^*}{{G_i}^*}a_ig_i \\[5pt]
\displaystyle \dot{\delta}&=& \displaystyle\frac{\kappa}{\tau_c} \left( \sqrt{\frac{{A_1}^*a_1}{{A_2}^*a_2}} + 
                                             \sqrt{\frac{{A_2}^*a_2}{{A_1}^*a_1}} \right) \sin(\delta)  -\delta_0 \\
\end{array}
\end{equation} 
where we have used the abbrieviating notation $L_i^*=L_i({A_i}^*)$.
In order to identify suitable scale parameters we consider the system in the steady state.
This yields 
\begin{equation}
\label{eqLaserInter}
\begin{array}{r c l}
\displaystyle 0 &=& \displaystyle e_i {G_i}^* + \frac{r_i {G_i}^*}{{A_i}^*}  
              + \frac{\kappa\sqrt{{A_1}^*{A_2}^*}}{{A_i}^*}\cos(\delta^*)-\frac{{L_i}^*}{{A_i}^*}\\[5pt]
\displaystyle 0 &=&\displaystyle \frac{p_i}{{G_i}^*}- s_i - r_i- 
            e_i {A_i}^* \\[5pt]
\displaystyle 0&=&\displaystyle \frac{\kappa}{\tau_c} \left( \sqrt{\frac{{A_1}^*}{{A_2}^*}} + 
                                 \sqrt{\frac{{A_2}^*}{{A_1}^*}} \right) \sin(\delta^*)  -\delta_0 \\
\end{array}
\end{equation} 
In the socio-economic model studied in the previous section we have seen that it is advantageous to 
group gain and loss terms together. In this way we obtain scale parameters that correspond to 
interpretable time scales. Grouping the terms in the laser model in this way we can identify the scale parameters
\begin{equation}
\label{eqLaserTimesScale}
\begin{array}{r c l}
\displaystyle \alpha_i &:=& \displaystyle e_i {G_i}^* + \frac{r_i {G_i}^*}{{A_i}^*} = - 
            \frac{\kappa\sqrt{{A_1}^*{A_2}^*}}{{A_i}^*}\cos(\delta^*)+\frac{{L_i}^*}{{A_i}^*}\\[5pt]
\displaystyle \beta_i &:=& \displaystyle \frac{p_i}{{G_i}^*} = s_i + r_i + e_i {A_i}^* 
\end{array}
\end{equation} 
Note that we have treated the coupling term in the intensity equation as a loss term. 
This decision is motivated by the fact that frequency locking in experiments is generally 
observed to be anti-phase \cite{Roy:Lasers}. In this case $\delta^*\approx\pi$ and $cos(\delta^*)<0$. The 
analysis presented below will show that it is generally reasonable to consider the 
coupling term as a loss term. Our mathematical treatment of the system does not depend critically
on the negativity of the coupling term. However, if the coupling term were positive 
it would be advisable to define the scale parameters in a different way for the sake of interpretation.  

The way in which the scale parameters have been defined above allows us to identify the 
parameter $\alpha_i$ as the characteristic timescale of the intensity equation if $\pi/2\leq \delta^* \leq 3\pi/2$.
In this case $\alpha_i$ is the per-capita (per-photon) loss rate of intensity of photons in the 
laser mode. In other words this $\alpha_i$ is the inverse of the typical lifetime of a photon 
in the laser mode of laser $i$. Likewise $\beta_i$ is the characteristic timescale of the gain equation 
or the inverse of the typical lifetime of atoms in the upper laser level.

In order to eliminate the constant factors from Eq.~(\ref{eqLaserLong}) we need to define additional 
scale variables which measure the contribution of the individual terms to the total gain and loss rates.
We define the additional scale parameters 
\begin{equation}
\label{eqLaserScaleFractions}
\begin{array}{r c l}
\displaystyle \chi_i&:=& \displaystyle \frac{1}{\alpha_i} \frac{{L_i}^*}{{A_i}^*},\\[5pt]
\displaystyle \sigma_i&:=&\displaystyle  \frac{1}{\beta_i} (r_i + e_i {A_i}^*), \\[5pt]
\displaystyle \gamma_i&:=&\displaystyle \frac{1}{\alpha_i} e_1{G_i}^*  
\end{array}
\end{equation}
and the complementary parameters
\begin{equation}
\label{eqLaserComplementFractions}
\begin{array}{r c l}
\displaystyle \tilde{\chi}_i&=& \displaystyle 1-\chi_i = -\frac{1}{\alpha_i} \frac{\kappa\sqrt{{A_1}^*{A_2}^*}}{{A_i}^*}cos(\delta^*), \\[5pt]
\displaystyle \tilde{\sigma}_i&=& \displaystyle 1-\sigma_i =  \frac{1}{\beta_i} s_i, \\[5pt]
\displaystyle \tilde{\gamma}_i&=& \displaystyle 1-\gamma_i =  \frac{1}{\alpha_i} \frac{r_i {G_i}^*}{{A_i}^*}.  
\end{array}
\end{equation}
Note that
\begin{equation}
\label{eqLaserScaleComment}
\displaystyle \gamma_i=\displaystyle \frac{1}{\beta_i\sigma_i} e_i {A_i}^*, 
\displaystyle \tilde{\gamma}_i=\displaystyle \frac{1}{\beta_i\sigma_i} r_i. 
\end{equation}
Finally we define
\begin{equation}
\rho:=\displaystyle \frac{A_1^*}{A_2^*}.   
\end{equation}
We can interpret the scale parameters as follows. The parameter $\chi_i$ denotes 
the fraction of the losses of intensity in laser $i$ that occur because of absorbtion or 
losses from the resonator. The complementary parameter $\tilde{\chi}_i$ is the fraction 
of the losses of laser $i$ that occur because of the coupling to the other laser.
The parameter $\sigma_i$ describes the fraction of the total loss of gain of laser $i$ that 
corresponds to the production of photons in the laser mode. In a similar way the parameter $\gamma_i$ 
denotes the contribution of stimulated emission to this fraction. In other words, $\gamma_i$ denotes the 
fraction of photons in the laser mode of laser $i$ that were created by spontaneous emission. 

The scale parameters allow us to write the model equations as
\begin{equation}
\label{eqLaserNorm}
\begin{array}{r c l}
\displaystyle \dot{a}_i&=& \displaystyle \alpha_i (\gamma_i  a_i g_i + \tilde{\gamma}_i g_i 
              - \tilde{\chi}_i \sqrt{a_1a_2} \frac{\cos(\delta)}{\cos(\delta^*)}-\chi_i l_i(a_i))\\[5pt]
\displaystyle \dot{g}_i&=&\displaystyle \beta_i ( 1 - \tilde{\sigma}_i g_i- \sigma_i( \tilde{\gamma}_i g_i + \gamma_i a_ig_i)) \\[5pt]
\displaystyle \dot{\delta}&=&\displaystyle \frac{\kappa}{\tau_c} \left( \sqrt{\frac{\rho a_1}{a_2}} + 
                                             \sqrt{\frac{a_2}{\rho a_1}} \right) \sin(\delta)  -\delta_0 \\
\end{array}
\end{equation} 

\subsection{Computation of the Jacobian}
Having completed the normalization we can now compute the Jacobian of the model. We start 
by defining the exponent parameter
\begin{equation} 
\label{eqLaserGradient}
\mu_i:=\displaystyle \left. \frac{\partial }{\partial a_i} l_i(a_i)\right|_{a_i=1}\\
\end{equation} 

Using this parameter we can write the Jacobian in the steady state as 
\begin{widetext}
{
\small
\begin{equation}
{\rm \bf J}=\left( \begin{array}{c c c c c} 
\alpha_1 & & & & \\
& \alpha_2 & & & \\
& & \beta_1 & & \\
& & & \beta_2 & \\
& & & & \kappa/\tau_c  \end{array} \right)
 \times 
\left( \begin{array} {c c c c c}
\gamma_1 -0.5 \tilde{\chi}_1 - \chi_1 \mu_1&  - 0.5 \tilde{\chi}_1 & 1 & 0 & \tilde{\chi}_1 \tan (\delta^*) \\
- 0.5 \tilde{\chi}_2 & \gamma_2-0.5\tilde{\chi}_2-\chi_2\mu_2 & 0 & 1 & \tilde{\chi}_2 \tan (\delta^*) \\
-\sigma_1\gamma_1 & 0 & - 1 & 0 & 0 \\
0 & -\sigma_2\gamma_2 & 0 & -1 & 0 \\
0.5(\rho^{1/2}-\rho^{-1/2})\sin(\delta^*) & 0.5(\rho^{-1/2}-\rho^{1/2})\sin(\delta^*) & 0 & 0
& (\rho^{1/2}+\rho^{-1/2})\cos(\delta^*)
\end{array} \right) 
\end{equation}
}
\end{widetext}
Let us assume that both lasers operate with same intensity
in the steady state. This implies $\rho=1$. In this case all elements of the fifth row of the Jacobian 
vanish except for $J_{5,5}$. Therefore we can immediately identify the eigenvalue 
$\lambda_5=J_{5,5}=2\cos(\delta^*)$. The existence of this eigenvalue proves that stable 
steady states are only possible for $\pi/2\leq\delta^*\leq 3\pi/2$. In this interval 
the stability of steady states is determined by the other four eigenvalues and therefore by the reduced
Jacobian
{
\begin{equation}
\begin{array}{c}
{\rm {\bf J}_{red}} = \left( \begin{array}{c c c c} 
\alpha_1 & & & \\
& \alpha_2 & & \\
& & \beta_1 & \\
& & & \beta_2 \end{array}\right) \times \\ 
\left( \begin{array} {c c c c}
\gamma_1 -0.5 \tilde{\chi}_1 - \chi_1 \mu_1&  - 0.5 \tilde{\chi}_1 & 1 & 0\\
- 0.5 \tilde{\chi}_2 & \gamma_2-0.5\tilde{\chi}_2-\chi_2\mu_2 & 0 & 1 \\
-\sigma_1\gamma_1 & 0 & - 1 & 0 \\
0 & -\sigma_2\gamma_2 & 0 & -1 
\end{array} \right)
\end{array}
\end{equation}
}
\subsection{Computation of Bifurcations}
\begin{figure}
\centering
\includegraphics[width=7.5cm]{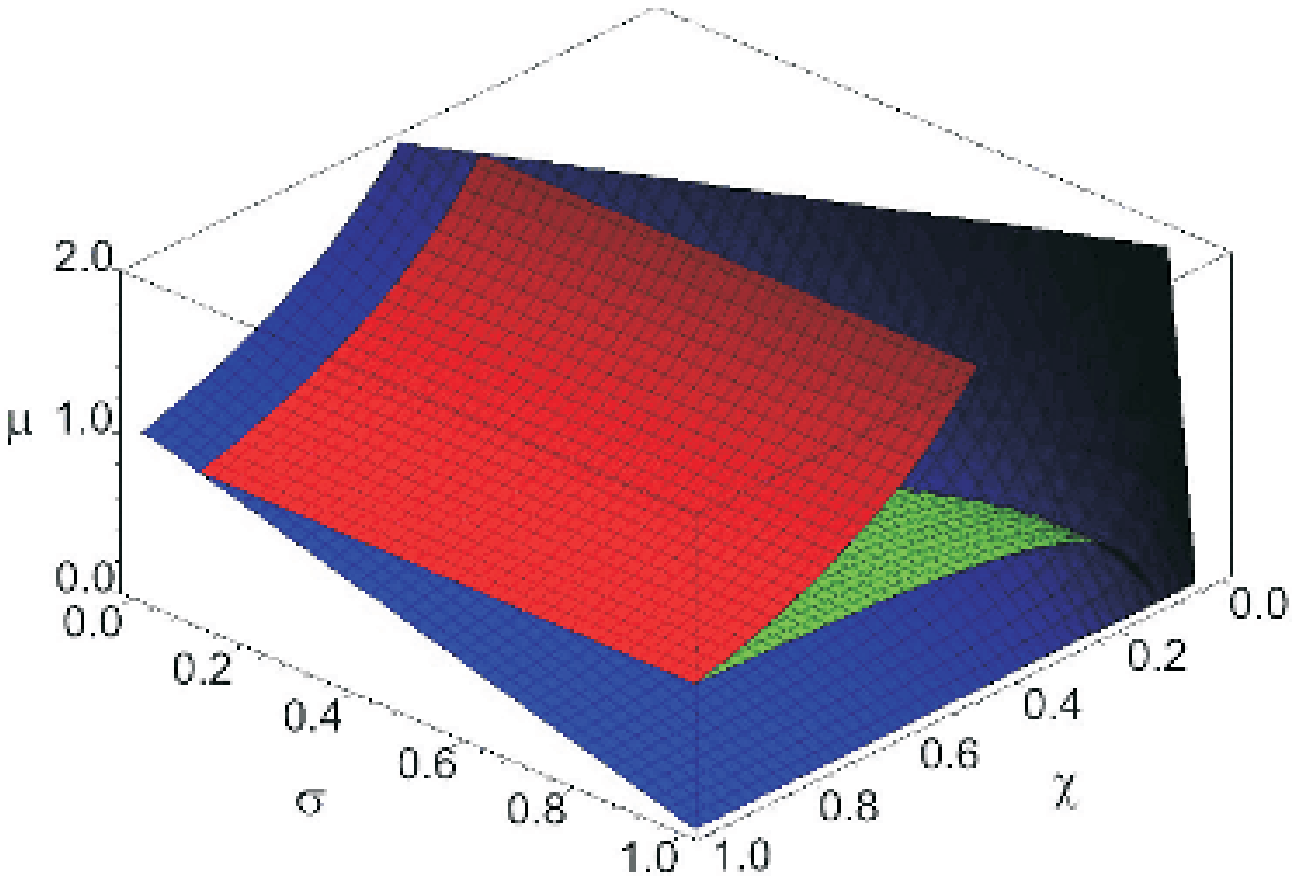}\\
\includegraphics[width=7.5cm]{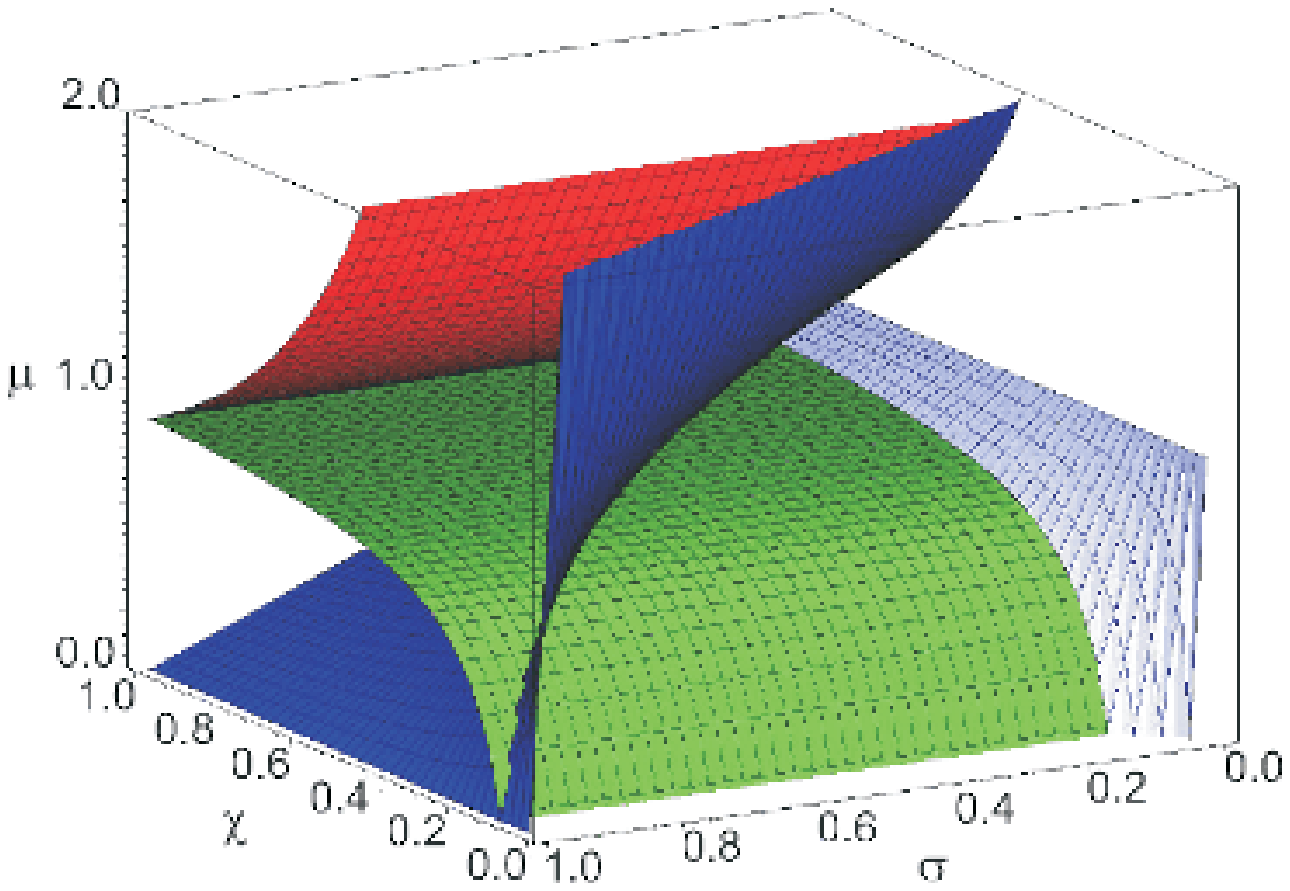}
\caption{(Color) Two different perspectives on a three parameter bifurcation diagram of the coupled laser 
system. The steady state under consideration is stable in the volume that is located at high values of
$\chi$ and $\mu$. The diagrams show the two saddle-node bifurcation surfaces S1 (dark blue)
and S2 (light blue) and the two Hopf bifurcation surfafaces H1 (red)
and H2 (green). At the intersection of these surfaces codimension-2 Takens-Bogdenov, 
Gavrilov-Guckenheimer and double Hopf bifurcations are formed. All codimension-2 lines meet in a codimension-4 
quadruple point bifurcation. While the codimension-1 indicate the borders of the region of local stability,
the bifurcations of higher codimension enable us to draw certain conclusions on the global dynamics.
\label{figLasers}}
\end{figure}

Let us now study the bifurcations of the generalized laser model based on the reduced Jacobian.
We simplify the reduced Jacobian further by assuming that the two lasers in the system 
are described by identical parameters
($\alpha_1=\alpha_2=:\alpha$, $\beta_1=\beta_2=:\beta$, $\gamma_1=\gamma_2=:\gamma$, 
$\chi_1=\chi_2=:\chi$, $\mu_1=\mu_2=:\mu$ and $\sigma_1=\sigma_2=:\sigma$). This assumption 
is valid for many coupled laser systems that are studied in experiments.
Furthermore we set $\alpha=1$ by means of time normalization. 
As in the previous example we apply the method of resultants to compute the local bifurcations
\cite{HopfPaper}. This reveals the bifurcation hypersurfaces of saddle-node type 
\begin{equation}
\begin{array}{r c c c l}
\displaystyle {\rm S1:} & \mu & = & \displaystyle \frac{\gamma(\sigma-1)}{\chi}        & \mbox{for } \sigma<\frac{b}{\gamma} \\[5pt]
\displaystyle {\rm S2:} & \mu & = & \displaystyle \frac{\gamma(\sigma-1)+1-\chi}{\chi} & \mbox{for } \sigma<\frac{b}{\gamma}
\end{array}
\end{equation}
and the Hopf bifurcation hyper-surfaces   
\begin{equation}
\begin{array}{r c c c l}
\displaystyle {\rm H1:} & \mu & = & \displaystyle \frac{\gamma-b}{\chi}        & \mbox{for } \sigma<\frac{b}{\gamma} \\[5pt]
\displaystyle {\rm H2:} & \mu & = & \displaystyle \frac{\gamma-b+\chi-1}{\chi} & \mbox{for } \sigma<\frac{b}{\gamma}
\end{array}
\end{equation}
These bifurcation surfaces are shown in Fig.~\ref{figLasers}. The steady state under consideration 
is stable in the top-most volume in the front of the left diagram. The right diagram shows the bifurcation 
surfaces from a different perspective so that the stable area is now in the back of the diagram.
The stability is lost if one of the Hopf or saddle-node bifurcation surfaces is crossed. 
Note, that for $\sigma<b/\gamma$ both Hopf bifurcation surfaces are independent of $\sigma$. 
At $\sigma=b/\gamma$ both surfaces end simultaneously in Takens-Bogdanov bifurcation lines.
The Takens-Bogdanov line of the Hopf bifurcation surface H1 is formed as it meets the 
bifurcation surface S1. At the same value of $\sigma$ the surface H2 ends in a 
Takens-Bogdanov bifurcation line as it meets S2. In addition to the Takens-Bogdanov lines 
there are two other codimension-2 bifurcation lines. A Gavrilov-Guckenheimer bifurcation 
is formed as H2 intersects S1. Finally, H1 and H2 meet in a double Hopf bifurcation line at $\chi=1$.
All of these codimension-2 bifurcation lines meet in a single point at $\chi=1$, $\sigma=b/\gamma$.
This point is a bifurcation point of codimension-4. 

As in the previous example the existence of bifurcations of higher codimension enables us to draw
conclusions on the global dynamics of the system. The presence of Takens-Bogdanov bifurcations 
proves that a homoclinic bifurcation has to exist. This bifurcation corresponds to the formation of 
a homoclinic loop that connects one saddle to itself. Homoclinic loops are known to cause bursting behavior.
In laser experiments such dynamics were observed in \cite{DeShazer:Lasers,Roy:Lasers}.
Furthermore homoclinic bifurcations play an important role in the formation of Shilnikov chaos 
\cite{Glendinning:StabilityInstabilityChaos}. In the previous section we have already mentioned 
that the Gavrilov-Guckenheimer bifurcation indicates the existence of a region of complex 
(at least quasiperiodic) dynamics. In a similar way the double Hopf 
bifurcation indicates the generic existence of a region in which chaos can be, at least transiently, 
observed \cite{Kuznetsov:Elements}. 
 
The codimension-2 bifurcation lines do not only show that certain global dynamics do generically exist 
in the system. They also point to the parameter region where such dynamics are likely. For instance 
chaotic dynamics is likely close to the double Hopf bifurcation line. In this way the relatively simple 
computation of codimension-2 bifurcations can be used to locate interesting regions in parameter space.
In these regions more time consuming techniques, like for instance the computation of Lyapunov exponents 
in specific models can then be applied.    

\section{A general food web model \label{secFoodWeb}}
In the previous sections we have applied the proposed approach to a generalized social model
and to a partially generalized model of coupled lasers. In this section the same procedure is applied to a generalized 
model, which describes the dynamics of ecological food webs. The long term behavior and in particular the stability of food webs 
is investigated in a large number of theoretical and observational studies \cite{Steele:Plankton,Ludwig:Budworm,Blasius:Synchronization}.
In the following we compute the Jacobian of a generalized food web model.
Since this model is more involved than the ones considered in the previous sections some additional difficulties arise. 
Nevertheless, the Jacobian can be constructed by following the algorithm outlined in Sec.~\ref{secLasers}.  
   
\subsection{Formulation of the ecological model}
Let us consider a food web of $N$ ecological populations. We denote the size of 
these populations with dynamical variables $X_1,\ldots,X_N$. For the sake of generality we do 
not distinguish between producer and consumer populations. Every population in the model can grow 
either by primary production (e.g. photosynthesis) or by consumption of prey (predation). The size of a given 
population can decrease because of predation by others or because of other causes of mortality 
(e.g. natural aging, diseases). We denote the primary production of population $n$ by the function $S_n(X_n)$. 
The growth of population $n$ by predation on others is described by the functional response $F_n(X_1,\ldots,X_N)$.
The loss inflicted by a higher predator $m$ is described by the function $L_{m,n}(X_1,\ldots,X_N)$.
The mortality of species $n$ is denoted as $M_n(X_n)$. Finally, we introduce the 
parameter $\eta$ which denotes the portion of consumed prey biomass that is converted into predator biomass. 
This yields the model
\begin{equation}
\label{eqFWModel}
\begin{array}{r c l} 
\displaystyle \dot{X}_n &\displaystyle =& \displaystyle S_n(X_n) + \eta_n F_n(X_1,\ldots,X_N) - M_n(X_n) \\[5pt]
\displaystyle           &\displaystyle  & \displaystyle - \sum_{m=1}^N L_{m,n}(X_1,\ldots,X_N) 
\end{array}
\end{equation}  
where $n=1,\ldots,N$. 

In order to describe natural food webs in a realistic way we have to take an additional constraint into account. 
The loss rate of population $n$ because of predation by population $m$ is not unrelated to the growth rate $F_m$ of 
population $m$. In order to quantify this relationship we introduce 
the auxilliary variable $T_m(X_1,\ldots,X_N)$ which denotes the total amount of prey that is available to species $m$. 
It is reasonable to write $T_m(X_1,\ldots,X_N)$ as 
\begin{equation}
\label{eqFoodSum}
T_m(X_1,\ldots,X_N)=\sum_{n=1}^N C_{m,n}(X_n),
\end{equation}
where $C_{m,n}(X_n)$ are auxilliary variables that describe the contribution of the population $n$
to the total amount of food available to population $m$. The value of $C_{m,n}(X_n)$ depends not only 
on the density of population $n$ but also the preference of the predators for specific prey and on the success rate of 
attacks \cite{Gentleman:Responses}.  

We can now write the amount of prey consumed by population $m$ as 
\begin{equation}
F_m(X_1,\ldots,X_N) = F_m(T_m,X_m).
\end{equation}
Since species $n$ contributes a fraction $C_{m,n}(X_n)/T_m(X_1,\ldots,X_N)$ to the total amount 
of prey available to species $m$ it is reasonable to assume that it contributes the same fraction 
to the consumption rate. That means,
\begin{equation}
\label{eqAux}
L_{m,n}(X_1,\ldots,X_N) = \frac{C_{m,n}(X_n)}{T_m(X_1,\ldots,X_N)} F_m(T_m,X_m)
\end{equation} 

\subsection{Normalization of the ecological model}
So far we have formulated a model of a $N$-species food web. In contrast to the models 
considered in the previous section we had to impose additional constaints in form of the 
algebraic equation Eq.~(\ref{eqAux}). 
In the following we apply our normalization procedure to normalize these algebraic equations
in addition to the ODEs.    

Like in the previous sections we use asterics to indicate the values of functions and variables 
in the steady state. Using this notation, we define normalized state variables
\begin{equation}
\label{eqDefVar}
x_n=\frac{X_n}{{X_n}^*} 
\end{equation}
normalized auxilliary variables
\begin{equation}
\begin{array}{r c l}
t_m&:=&\displaystyle \frac{T_m({X_1}^*x_1,\ldots,{X_N}^*x_N)}{{T_m}^*}, \\[5pt]
c_{m,n}&:=&\displaystyle \frac{C_{m,n}({X_n}^*x_n)}{{C_{m,n}}^*}
\end{array}
\end{equation}
and normalized functions
\begin{equation}
\label{eqDefFunc}
\begin{array}{r c l}
s_n(x_n)&:=&\displaystyle \frac{S_n({X_n}^*x_n)}{{S_n}^*},\\[5pt]
m_n(x_n)&:=&\displaystyle \frac{M_n({X_n}^*x_n)}{{M_n}^*},\\[5pt]
f_n(x_1,\ldots,x_N)&:=&\displaystyle \frac{F_n({X_1}^*x_1,\ldots,{X_N}^*x_N)}{{F_n}^*},\\[5pt]
l_{m,n}(x_1,\ldots,x_N)&:=&\displaystyle \frac{L_{m,n}({X_1}^*x_1,\ldots,{X_N}^*x_N)}{{L_{m,n}}^*}.
\end{array}
\end{equation}
 
As the second step of the normalization we rewrite the model in terms of the normalized 
quantities. We start by substituting the definitions Eqs.~(\ref{eqDefVar})-(\ref{eqDefFunc}) 
into Eq.~(\ref{eqFWModel}). This yields
\begin{equation} 
\label{eqInterModel}
\begin{array}{r c l}
\dot{x}_n & = & \displaystyle 
\frac{{S_n}^*}{{X_n}^*} s_n(x_n) 
+ \frac{\eta_n{F_n}^*}{{X_n}^*} f_n(x_1,\ldots,x_N) \\[5pt] 
& & \displaystyle - \frac{M_n^*}{{X_n}^*}m_n(x_n) 
- \sum_{m=1}^N \frac{{L_{m,n}}^*}{{X_n}^*}l_{m,n}(x_1,\ldots,x_N)
\end{array} 
\end{equation}  
In the next step we have to define suitable scale variables. 
We start with the time scales  
\begin{equation}
\label{eqDefAlpha}
\alpha_n := \frac{{S_n}^*}{{X_n}^*} + \frac{\eta_n{F_n}^*}{{X_n}^*}  
= \frac{M_n^*}{{X_n}^*} + \sum_{m=1}^N \frac{{L_{m,n}}^*}{{X_n}^*}.
\end{equation} 
These scale parameters quantify the rate of biomass flow in the steady state. 
The relative contributions of the different processes to the biomass flow can be described by
\begin{equation}
\begin{array}{r c l}
\rho_n&:=&\displaystyle \frac{1}{\alpha_n}\frac{\eta_n{F_n}^*}{{X_n}^*},\\[5pt]
\tilde{\rho}_n&:=&\displaystyle 1-\rho_n=\frac{1}{\alpha_n}\frac{{S_n}^*}{{X_n}^*},\\[5pt]
\sigma_n&:=&\displaystyle \frac{1}{\alpha_n} \sum_{m=1}^N \frac{{L_{m,n}}^*}{{X_n}^*},\\[5pt]
\tilde{\sigma}_n&:=&\displaystyle 1-\sigma_n=\frac{1}{\alpha_n}\frac{M_n^*}{{X_n}^*},\\[5pt]
\beta_{m,n}&:=&\displaystyle \frac{1}{\alpha_n\sigma_n} \frac{{L_{m,n}}^*}{{X_n}^*}.
\end{array}
\end{equation}

Before we discuss the interpretation of these parameters in more detail let us 
examine the algebraic equations in the model. We start with Eq.~(\ref{eqAux}). 
In this case the substitution of the normalized quantities yields
\begin{equation}
\begin{array}{r c l}
\label{eqAuxInt}
l_{m,n}(x_1,\ldots,x_N) & =& \displaystyle \frac{{C_{m,n}}^*{F_m}^*}{{T_m}^*{L_{m,n}^*}} \frac{c_{m,n}}{t_m} f_m(t_m,x_m)
\\ &=& \displaystyle \frac{c_{m,n}}{t_m} f_m(t_m,x_m),
\end{array}
\end{equation} 
where we have used Eq.~(\ref{eqAux}) in the second step.  

The last equation which we have to consider is Eq.~(\ref{eqFoodSum}). In this case we obtain
\begin{equation}
\label{eqFoodSumInt}
t_m= \sum_{n=1}^N \frac{{C_{m,n}}^*}{{T_m}^*} c_{m,n}.
\end{equation}
In contrast to Eq.~(\ref{eqAuxInt}) the constant factors do not cancel out in this equation. 
It is therefore useful to define the additional scale parameters 
\begin{equation}
\chi_{m,n}=\frac{{C_{m,n}}^*}{{T_m}^*}.
\end{equation} 
We can now write the food web model as 
\begin{equation}
\label{eqNormModel}
\begin{array}{r c l}
\dot{x}_n &=&\alpha_n(\tilde{\rho}_n s_n(x_n) + \rho_n f_n(t_n,x_n) - \tilde{\sigma_n}m_n(x_n) \\ 
          & &  - \sigma_n\sum_{m=1}^N \beta_{m,n}l_{m,n}(x_1,\ldots,x_N))
\end{array} 
\end{equation}  
for $n=1\ldots N$ and
\begin{equation}
\label{eqNormAux}
l_{m,n}(x_1,\ldots,x_N) = \frac{c_{m,n}}{t_m} f_m(t_m,x_m),
\end{equation}
\begin{equation}
\label{eqFoodSumNorm}
t_m= \sum_{n=1}^N \chi_{m,n} c_{m,n}.
\end{equation}

As a result of the normalization we have identified a set of scale parameters. Let us now discuss these 
parameters in more detail.

Based on our experience from the previous section we can identify the parameters $\alpha_1,\ldots,\alpha_N$
immediately as the characteristic timescales of the system. The parameter $\alpha_n$ denotes the per-capita
birth and death rate of population $n$ in the steady state. In other words, $\alpha_n$ is the inverse of the 
typical lifetime of individuals from population $n$. 

The parameter $\rho_n$ is defined as the per-capita biomass gain of population $n$ divided by 
the total per-capita growth rate $\alpha_n$. In other words, this parameter describes what 
fraction of the growth rate of population $n$ is gained by predation. The complementary 
parameter $\tilde{\rho}_n$ denotes the fraction of the growth that originates from primary production. 
If population $n$ consists of primary producers the corresponding parameter values 
are $\rho_n=0, \tilde{\rho}_n=1$. A population of consumers has $\rho_n=1, \tilde{\rho}_n=0$.
Some organisms are known which are capable of primary production and predation, for these $\rho_n$ and $\tilde{\rho}_n$
can have fractional values between zero and one.

The parameters $\sigma_n$ and $\tilde{\sigma}_n$ are very similar to $\rho_n$ and $\tilde{\rho}_n$.
While $\rho_n$ and $\tilde{\rho}_n$ describe the relativ importance of the growth terms, $\sigma_n$ and $\tilde{\sigma}_n$
denote the relative importance of loss terms. The parameter $\sigma_n$ is defined as the fraction of the total loss
rate of population $n$ that occurs because of predation by explicitly modelled predators. The fraction of losses that is 
caused by all other sources of mortality (e.g. natural mortality, diseases, predation by predators 
that are not explicitly modeled) is given by $\tilde{\sigma}_n$. We can therefore say, that the value 
of $\sigma_n$ indicates the predation pressure on species $n$.

The parameter $\beta_{m,n}$ is defined as the per-capita loss rate of population $n$ because of predation by population
$m$ divided by the total loss rate of population $n$ because of predation $\alpha_n\sigma_n$. This means that
$\beta_{m,n}$ denotes the relative weight of the contribution of species $m$ to predative loss rate of 
species $n$. If, for example, population $1$ is the only population that preys upon population $2$ then 
the corresponding parameter is $\beta_{1,2}=1$. However, if population $1$ causes only 50\% of the 
losses while the other half is caused by predation by cannibalistic predation of population $2$ upon itself
then we find $\beta_{1,2}=0.5$, $\beta_{2,2}=0.5$.   

Finally, the parameter $\chi_{m,n}$ measures the relative contribution of population $n$ to the total amount
of food that is available to species $m$. Note that, $\chi_{n,m}\neq\beta_{n,m}$. For example predation 
by species $1$ may only contribute, say $1/10$th of the total predative loss of species $2$. But, at the same 
time species $1$ may be the only species which is consumed by species $2$. In this situation we have $\beta_{1,2}=0.1$
but $\chi_{1,2}=1$.   

\subsection{Computation of the Jacobian}
Like in the previous examples we start the computation of the Jacobian by defining 
a set of exponent parameters that describe the nonlinearity of the ecological 
processes in the steady state
\begin{equation}
\begin{array}{r c l}
\phi_n         &:=& \left.\frac{\partial}{\partial x_n} s_n(x_n)         \right|_{{\rm \bf x}={\rm \bf x^*}}, \\[5pt] 
\mu_n          &:=& \left.\frac{\partial}{\partial x_n} m_n(x_n)         \right|_{{\rm \bf x}={\rm \bf x^*}}, \\[5pt]
\lambda_{m,n}  &:=& \left.\frac{\partial}{\partial x_n} c_{m,n}(x_n)     \right|_{{\rm \bf x}={\rm \bf x^*}}, \\[5pt]
\gamma_n       &:=& \left.\frac{\partial}{\partial t_n} f_n(t_n,x_n)     \right|_{{\rm \bf x}={\rm \bf x^*}}, \\[5pt]
\psi_n         &:=& \left.\frac{\partial}{\partial x_n} f_n({t_n}^*,x_n) \right|_{{\rm \bf x}={\rm \bf x^*}}. 
\end{array}
\end{equation} 
To a large extend these definitions follow the same scheme that we have applied in the previous sections.
Note however that ${t_n}^*$ appears in the definition of $\psi_n$. In order to understand the reason 
for this definition we have to be aware of the fact that cannibalism can occur. Consequently, the 
amount of food $t_n$ that is available to the predator $x_n$ can depend on $x_n$ itself.
The definition given above ensures that $\psi_n$ denotes only the nonlinearity of the 
predator-dependence of the response function. 

The derivation of the Jacobian for the general food web model is analogous to the previous examples.
For this reason the calculation is presented in appendix \ref{AppA}. As a result we find that the 
non-diagonal elements of the Jacobian can be written as  
\begin{equation}
\label{eqFoodJacobian1}
\begin{array}{r c l}
J_{n,i}&=&\alpha_n (\rho_n\gamma_n\chi_{n,i}\lambda_{n,i} -\sigma_n(\beta_{i,n}\psi_i\\
       & &+\sum_{m=1}^N \beta_{m,n}
\lambda_{m,i}(\gamma_m-1)\chi_{m,i}))
\end{array}
\end{equation}
and the diagonal elements as
\begin{equation}
\label{eqFoodJacobian2}
\begin{array}{r c l}
J_{i,i}&=&\alpha_i (\tilde{\rho}_i\phi_i 
+\rho_i(\gamma_i\chi_{i,i}\lambda_{i,i}+\psi_i) \\
& & -\tilde{\sigma}_i \mu_i 
-\sigma_i(\beta_{i,i}\psi_i+\\
& & +\sum_{m=1}^N \beta_{m,i}\lambda_{m,i}((\gamma_m-1)\chi_{m,i}+1)))  
\end{array}
\end{equation}

The parameter $\phi_n$ that appears in these equations describes the nonlinearity of the primary 
production of species $n$. In an environment where nutrient are abundant and no other limiting factors 
exist, it is reasonable to assume that the primary production $S_n$ is proportional to the number of 
primary producers $X_n$. This corresponds to $\sigma_n=1$. On the other hand, if nutrients are scarce 
then the primary production is limited by the nutrient supply. In this case the primary production 
is a constant function of the number of primary producers and $\sigma_n=0$. In a realistic 
situation $\sigma_n$ has a value between $0$ and $1$. This value can be interpreted as an 
indicator of the nutrient availability in the system. 

The parameter $\mu_n$ denotes the nonlinearity of the mortality rate of population $n$.
For top predators this parameter is often called \emph{exponent of closure} in the ecological literature.
More generally speaking, we call $\mu_n$ exponent of mortality. 
In most food web and food chain models this parameter is assumed to be either one or two.
However, it has been shown that fractional values between one and two may describe natural 
systems in a more realistic way. This is discussed in detail in \cite{Edwards:Exponent}.

The parameter $\gamma_n$ denotes the nonlinearity of the predation rate 
of population $n$ with respect to the prey density. If the prey is abundant
then the predation rate is not limited by prey availability and $\gamma_n=0$.
However, in most natural systems prey is relatively scarce. In this case the value 
of $\gamma_n$ can depend on the feeding strategy of population $n$. In the limit 
of scarce prey the value of $\gamma_n$ approaches one for predators of Holling type II
and up to two for predators of Holling type III. This is discussed in more detail in 
\cite{EnrichmentPaper}.
     
The parameter $\psi_n$ describes the dependence of the predation rate of population $n$
on the population density of $n$ itself. In most models it is assumed that predators of the same species 
feed independently without direct interference. In this case the predation rate is a linear
function of the predator density and $\psi_n=1$. However, intraspecific competition or social 
interactions can introduce nonlinearities in the predator dependence of the predation rate.
If these effects are taken into account we find $0<\psi_n<1$. An extreme example are the so-called 
ratio dependent models \cite{Abrams:Debate}. These models correspond to $\psi_n=1-\gamma_n$.

Finally, the parameter $\lambda_{m,n}$ describes the nonlinearity of the contribution of 
population $m$ to the diet of population $n$. In the simplest case we can assume that 
the predator does not distuinguish actively between different prey types. An example 
of such a predator is an aquatic filtration feeder. In this case the contribution of a 
suitable prey population $m$ is proportional to the size of the population $m$ and 
$\lambda_{m,n}=1$.
However, other predators attack their prey individually. These predators 
may adapt their strategies to a specific prey population if they have sufficient practice.
In this case the success rate of attacks as well as the enounter rate increase linearly with
the density of the prey population. In effect that leads to a quadratic dependence and therefore 
$\lambda_{m,n}=2$. Another type of behavior 
is exhibited by predators that need to choose their prey actively in order to obtain all nutrients 
that are necessary for growth. As an extreme example we could imagine a predator that requires 
exactly the right composition of its diet. In this case we find $\lambda_{m,n}=0$ for all prey populations
$m$ but the limiting one.   
 
\subsection{Stability of generalized food webs}
In Sec.~\ref{secSocial} and Sec.~\ref{secLasers} we have used bifurcation 
analysis to study the Jacobians of the normalized models. For the investigation of food webs this 
analysis is also possible. In fact, bifurcation analysis has been used in \cite{HopfPaper,EnrichmentPaper,
ExponentPaper} to study the dynamics of generalized food chains. The generalized food chains considered 
in these papers are particular examples of the more general class of food webs considered here. 
The bifurcation analysis of other examples of generalized food webs can reveal new ecological insights. 
For instance we can consider bifurcation diagrams that correspond to different 
food web geometries. In this way one can investigate which food webs are characterized by a high 
local stability. Furthermore we can investigate the impact of certain ecological mechanisms 
on the stability or use higher codimension bifurcations to draw conclusions on the existence 
of complex dynamics. However, from the methodological point of view, these investigations 
are to a large extent parallel to the previous examples. Let us therefore describe a different way 
in which the derived Jacobian can be applied.

In order to investigate the stability of ecological systems random matrix models are frequently considered
\cite{May:ComplexityStability}. In this context the Jacobian of a food web is modeled 
by a random matrix. The eigenvalues of a large set of these randomly created matrices are then computed. 
In this way one can for instance obtain a relationship between the stability of a random food web and the 
connectivity (often measured in terms of the number of non-diagonal elements of the matrix). While these 
analyses have provided many interesting results it has often been argued that not all of the randomly 
created matrices are biologically reasonable \cite{Haydon:Stability}.

The Jacobian of a generalized food web of given size can be studied in a similar way. By choosing the scale 
and exponent parameters randomly from suitable distributions one can create a set of random but realistic Jacobians.
These Jacobians can then be studied by the numerical computation of eigenvalues. The advantage of this approach 
is that random matrices with certain ecological properties can be created. In this way one can for instance 
investigate the impact of ecological effects like intraspecific competition or cannibalism on the stability of the 
system. Although these investigations are certainly promising they require a much more detailed analysis and are 
therefore beyond the scope of the current paper.

\section{Discussion \label{secConclusions}}
In this paper we propose an approach to the investigation of generalized models which can be formulated
in form of balance equations. In these equations gain and loss terms determine the time evolution of the 
state variables of the system under consideration. The proposed normalization procedure enables the 
researcher to consider a system from an abstract point of view. This abstractness is caused by the way in which 
variables and functions are normalized. The normalization procedure separates the 
necessary information about the steady state (the scale parameters) from the information 
about the nonlinearities in the model (the exponent parameters). In this way it becomes 
possible to study the effect of the nonlinearities independently from the location of the 
steady state. Apart from enabling us to investigate generalized models, this separation 
of scale and exponent variables has other advantages. For instance the elements of 
the Jacobians that are derived in this way are generally relatively simple. More importantly, 
it is in almost all models possible to identify the parameters in such a way that they 
are easily interpretable in the context of the model. In many cases the general parameters are more directly accessible 
to measurements than the specific parameters that are normally used in modeling.
In \cite{EnrichmentPaper} it has been shown that the identification 
of the general parameters can in itself provide new insights.  

The basic idea behind the proposed approach is to consider the bifurcations as functions of 
a new set of parameters which only encode as much information as is needed to construct the Jacobian.
In this way we avoid the computation of steady states, which is in many cases more
complicated than the computation of bifurcations once the Jacobian is known. 
This enables us to compute the local bifurcation surfaces analytically (using computer algebra systems)
as a function of the scale and exponent parameters. As a result one is able to investigate 
the role of the nonlinearities in the gain and loss terms independently from the magnitude of these 
terms. In other words, one can study, with a high degree of generality, which functional forms 
of the gain and loss terms yields which dynamical behavior. Thus, this approach provides a bridge 
between modeling and stability analysis in dynamical systems theory. 

The focus of the bifurcation analysis is the computation of local codimension-1 bifurcation of the 
normalized steady state. But, as a byproduct of this analysis we obtain the local bifurcations of higher 
codimension, which can reveal many insights about the dynamics in the neighborhood of these bifurcations.   
For instance the presence of a codimension-2 double Hopf bifurcation is an indicator for chaotic 
dynamics. Although our analysis is purely local, it can therefore reveal insights on the global dynamics of the 
system. In particular it can indicate regions in parameter space where complex dynamics can be expected.   
This insight can be used to prove the generic existence of chaotic parameter regions in generalized 
models or to locate interesting regions for subsequent numerical investigation.

At present there are only few results on the implications of local bifurcations of codimension 
larger than two. However, the opinion has often been expressed that this line of mathematical research would benefit 
from more examples from physical applications \cite{Guckenheimer:DynamicalSystems}. The approach proposed here 
provides physicist with a tool for finding these examples. In return, new mathematical results on the implications of bifurcations 
of higher codimension can be applied directly to the investigation of generalized models of physical systems. 
In this way, the investigation of generalized models may lead to more efficient transport of mathematical 
discoveries into physical applications and back. 

Our approach to the investigation of generalized models is purely determistic. 
Noise and stochastic environmental fluctuations can not be accomodated 
adequately in the models considered here. 
However, the aim of the proposed approach is not to 
predict the time evolution of a given system, but rather to gain 
a general understanding of the functioning of a class of models. 
For systems with a low or moderate level of noise, the investigation of deterministic 
generalized models can still be useful, since it is unlikely that the presense of noise 
alters the local bifurcation structure qualitatively in this case.    

The abstract nature of the proposed analysis becomes apparent if one considers that we 
did not have to specify which steady state we consider. In a system in which multiple steady 
states exist the derived bifurcation diagram does therefore describe all of these steady states 
simulateously. This insight can for instance be used to prove general statements of the type: ``There can not 
be a stable steady state with \ldots''. 

In this paper, the main tool for the investigation of the derived Jacobians was bifurcation analysis. 
However, we have already remarked in our final example that the Jacobians can also 
be used in other ways. In particular large systems may contain too many parameters to be analysed efficiently 
with bifurcation diagrams. For these systems our method can be used in the random-matrix-like context, which is 
described in Sec.~\ref{secFoodWeb}. In this light, the proposed method can be seen as a link between the very 
general random matrix approach and specific models.   

Finally, let us remark that the approach presented here is not restricted to systems of ordinary differential
equations. The algorithm which we have outlined in Sec.~\ref{secLasers} can be applied in the same way to compute
the bifurcations of discrete time maps. Moreover, it can be used in certain systems of partial differential equations 
to compute local bifurcations of homogenous solutions including Turing bifurcations \cite{spatialpaper}. 

\begin{acknowledgments}
We would like to thank Wolfgang Ebenh\"oh who inspired this line of research by proposing  
a simple generalized food chain model.
This work was sup\-por\-ted by the Deu\-t\-sche For\-schung\-s\-ge\-mein\-schaft (FE 359/6).
\end{acknowledgments}

\appendix*
\section{Computation of the Jacobian of the general food web \label{AppA}}
In this appendix we present the explicit comptation of the Jacobian of the food web model.
We start by considering
\begin{equation}
\begin{array}{c}
\left. \frac{\partial}{\partial x_i} t_m(x_1,\ldots,x_N) \right|_{{\rm \bf x}={\rm \bf x^*}}\\ 
  =  \left. \frac{\partial}{\partial x_i} \sum_{n=1}^N \chi_{m,n} c_{m,n}(x_n) \right|_{{\rm \bf x}={\rm \bf x^*}}.
\end{array} 
\end{equation}
We write the derivative of the predation rate as
\begin{equation}
\begin{array}{r c l}
\frac{\partial}{\partial x_i} f_n(t_n,x_n) &=& 
\left( \frac{\partial}{\partial t_n} f_n(t_n,x_n) \right)\left( \frac{\partial}{\partial x_i} t_n \right) \\
& & + \left( \frac{\partial}{\partial x_n} f_n({t_n}^*,x_n) \right)\left( \frac{\partial}{\partial x_i} x_n \right).
\end{array}
\end{equation}
This yields
\begin{equation}
\left. \frac{\partial}{\partial x_i} f_n(t_n,x_n) \right|_{{\rm \bf x}={\rm \bf x^*}} = 
\left\{ 
\begin{array}{l l}
\gamma_n \chi_{n,i} \lambda_{n,i}           & \mbox{for }i\neq n \\
\gamma_n \chi_{n,i} \lambda_{n,i} + \psi_i  & \mbox{for }i = n \\
\end{array}
\right. 
\end{equation}
Using this result we compute 
\begin{equation}
\label{eqNormAuxWeb}
\begin{array}{r c l}
\left. \frac{\partial}{\partial x_i} l_{m,n}(x_1,\ldots,x_N) \right|_{{\rm \bf x}={\rm \bf x^*}} 
&=&  - \chi_{m,i}\lambda_{m,i}  + \gamma_m \chi_{m,i} \lambda_{m,i} + Q_1\\   
&=&  (\gamma_m-1)\chi_{m,i}\lambda_{m,i}+Q_1.
\end{array}
\end{equation}
where
\begin{equation}
Q_1=\left\{ \begin{array}{l l}
0 & \hspace{1cm}\mbox{for } n\neq i, m\neq i \\
\lambda_{m,i} & \hspace{1cm}\mbox{for } n=i, m\neq i \\
\psi_i  & \hspace{1cm}\mbox{for } n\neq i, m=i \\
\lambda_{m,i} +\psi_i & \hspace{1cm}\mbox{for } n=i, m=i \end{array}
\right.
\end{equation}
Let us now compute the derivatives of the sum in Eq.~(\ref{eqNormModel}).
We obtain
\begin{equation}
\begin{array}{c}
\left. \frac{\partial}{\partial x_i} \sum_{m=1}^N \beta_{m,n}l_{m,n}(x_1,\ldots,x_N)) 
\right|_{{\rm \bf x}={\rm \bf x^*}}\\ = \beta_{i,n}\psi_i+\sum_{m=1}^N \beta_{m,n}\lambda_{m,i}((\gamma_m-1)\chi_{m,i}+Q_2). 
\end{array}
\end{equation}
where 
\begin{equation}
Q_2=\left\{\begin{array}{l l} 
1 & \hspace{1cm}\mbox{for } i=n\\
0 & \hspace{1cm}\mbox{for } i\neq n \end{array} \right.
\end{equation}
Using this result we can now write the non-diagonal elements as 
\begin{equation}
\begin{array}{r c l}
J_{n,i}&=&\alpha_n (\rho_n\gamma_n\chi_{n,i}\lambda_{n,i} -\sigma_n(\beta_{i,n}\psi_i\\
& &+\sum_{m=1}^N \beta_{m,n}
\lambda_{m,i}(\gamma_m-1)\chi_{m,i})
\end{array}
\end{equation}
and the diagonal elements as
\begin{equation}
\begin{array}{r c l}
J_{i,i}&=&\alpha_i (\tilde{\rho}_i\phi_i 
+\rho_i(\gamma_i\chi_{i,i}\lambda_{i,i}+\psi_i)\\
& & -\tilde{\sigma}_i \mu_i 
-\sigma_i(\beta_{i,i}\psi_i\\
& &+\sum_{m=1}^N \beta_{m,i}\lambda_{m,i}((\gamma_m-1)\chi_{m,i}+1)))  
\end{array}
\end{equation}


\begin{thebibliography}{27}
\expandafter\ifx\csname natexlab\endcsname\relax\def\natexlab#1{#1}\fi
\expandafter\ifx\csname bibnamefont\endcsname\relax
  \def\bibnamefont#1{#1}\fi
\expandafter\ifx\csname bibfnamefont\endcsname\relax
  \def\bibfnamefont#1{#1}\fi
\expandafter\ifx\csname citenamefont\endcsname\relax
  \def\citenamefont#1{#1}\fi
\expandafter\ifx\csname url\endcsname\relax
  \def\url#1{\texttt{#1}}\fi
\expandafter\ifx\csname urlprefix\endcsname\relax\def\urlprefix{URL }\fi
\providecommand{\bibinfo}[2]{#2}
\providecommand{\eprint}[2][]{\url{#2}}

\bibitem[{\citenamefont{Holling}(1959)}]{Holling:Characteristics}
\bibinfo{author}{\bibfnamefont{C.~S.} \bibnamefont{Holling}},
  \bibinfo{journal}{The Canadian Entomologist} \textbf{\bibinfo{volume}{91}},
  \bibinfo{pages}{385} (\bibinfo{year}{1959}).

\bibitem[{\citenamefont{Gross et~al.}(2004)\citenamefont{Gross, Ebenh\"oh, and
  Feudel}}]{EnrichmentPaper}
\bibinfo{author}{\bibfnamefont{T.}~\bibnamefont{Gross}},
  \bibinfo{author}{\bibfnamefont{W.}~\bibnamefont{Ebenh\"oh}},
  \bibnamefont{and} \bibinfo{author}{\bibfnamefont{U.}~\bibnamefont{Feudel}},
  \bibinfo{journal}{Journal of theoretical biology}
  \textbf{\bibinfo{volume}{227}}, \bibinfo{pages}{349} (\bibinfo{year}{2004}).

\bibitem[{\citenamefont{Fussmann and Blasius}(2005)}]{Bernd:Sensitivity}
\bibinfo{author}{\bibfnamefont{G.}~\bibnamefont{Fussmann}} \bibnamefont{and}
  \bibinfo{author}{\bibfnamefont{B.}~\bibnamefont{Blasius}},
  \bibinfo{journal}{Biology Letters} \textbf{\bibinfo{volume}{1}},
  \bibinfo{pages}{9} (\bibinfo{year}{2005}).

\bibitem[{\citenamefont{Gross et~al.}(2005)\citenamefont{Gross, Ebenh\"{o}h,
  and Feudel}}]{ExponentPaper}
\bibinfo{author}{\bibfnamefont{T.}~\bibnamefont{Gross}},
  \bibinfo{author}{\bibfnamefont{W.}~\bibnamefont{Ebenh\"{o}h}},
  \bibnamefont{and} \bibinfo{author}{\bibfnamefont{U.}~\bibnamefont{Feudel}},
  \bibinfo{journal}{Oikos} \textbf{\bibinfo{volume}{109}}, \bibinfo{pages}{135}
  (\bibinfo{year}{2005}).

\bibitem[{\citenamefont{Usher}(1989)}]{Usher:DynasticCycles}
\bibinfo{author}{\bibfnamefont{D.}~\bibnamefont{Usher}},
  \bibinfo{journal}{Am.~Econ.~Rev.} \textbf{\bibinfo{volume}{79}},
  \bibinfo{pages}{1031} (\bibinfo{year}{1989}).

\bibitem[{\citenamefont{Feichtinger et~al.}(1974)\citenamefont{Feichtinger,
  Forst, and Piccardi}}]{Feichtinger:DynasticCycles}
\bibinfo{author}{\bibfnamefont{G.}~\bibnamefont{Feichtinger}},
  \bibinfo{author}{\bibfnamefont{C.~V.} \bibnamefont{Forst}}, \bibnamefont{and}
  \bibinfo{author}{\bibfnamefont{C.}~\bibnamefont{Piccardi}},
  \bibinfo{journal}{Chaos, Solitons \& Fractals} \textbf{\bibinfo{volume}{7}},
  \bibinfo{pages}{257} (\bibinfo{year}{1996}).

\bibitem[{\citenamefont{Guckenheimer and
  Holmes}(2002)}]{Guckenheimer:DynamicalSystems}
\bibinfo{author}{\bibfnamefont{J.}~\bibnamefont{Guckenheimer}}
  \bibnamefont{and} \bibinfo{author}{\bibfnamefont{P.}~\bibnamefont{Holmes}},
  \emph{\bibinfo{title}{Nonlinear oscillations, dynamical systems, and
  bifurcations of vector fields}} (\bibinfo{publisher}{Springer},
  \bibinfo{address}{Berlin}, \bibinfo{year}{2002}), \bibinfo{edition}{7th} ed.

\bibitem[{\citenamefont{Gross}(2004)}]{Thesis}
\bibinfo{author}{\bibfnamefont{T.}~\bibnamefont{Gross}},
  \emph{\bibinfo{title}{Population dynamics: General results from local
  analysis}} (\bibinfo{publisher}{Der Andere Verlag},
  \bibinfo{address}{T\"onning}, \bibinfo{year}{2004}).

\bibitem[{\citenamefont{Kuznetsov}(1995)}]{Kuznetsov:Elements}
\bibinfo{author}{\bibfnamefont{Y.~A.} \bibnamefont{Kuznetsov}},
  \emph{\bibinfo{title}{Elements of Applied Bifurcation Theory}}
  (\bibinfo{publisher}{Springer}, \bibinfo{address}{Berlin},
  \bibinfo{year}{1995}).

\bibitem[{\citenamefont{Gross and Feudel}(2004)}]{HopfPaper}
\bibinfo{author}{\bibfnamefont{T.}~\bibnamefont{Gross}} \bibnamefont{and}
  \bibinfo{author}{\bibfnamefont{U.}~\bibnamefont{Feudel}},
  \bibinfo{journal}{Physica D} \textbf{\bibinfo{volume}{195}},
  \bibinfo{pages}{292} (\bibinfo{year}{2004}).

\bibitem[{\citenamefont{Haken}(1983)}]{Haken:Lasers}
\bibinfo{author}{\bibfnamefont{H.}~\bibnamefont{Haken}},
  \emph{\bibinfo{title}{Synergetics: An Introduction}}
  (\bibinfo{publisher}{Springer Verlag}, \bibinfo{address}{Berlin},
  \bibinfo{year}{1983}).

\bibitem[{\citenamefont{Mandel}(1997)}]{Mandel:Lasers}
\bibinfo{author}{\bibfnamefont{P.}~\bibnamefont{Mandel}},
  \emph{\bibinfo{title}{Theoretical problems in cavity nonlinear optics}}
  (\bibinfo{publisher}{Cambridge University Press},
  \bibinfo{address}{Cambridge}, \bibinfo{year}{1997}).

\bibitem[{\citenamefont{Lauterborn and Kurz}(2005)}]{Lauterborn:Lasers}
\bibinfo{author}{\bibfnamefont{W.}~\bibnamefont{Lauterborn}} \bibnamefont{and}
  \bibinfo{author}{\bibfnamefont{T.}~\bibnamefont{Kurz}},
  \emph{\bibinfo{title}{Coherent Optics}} (\bibinfo{publisher}{Springer
  Verlag}, \bibinfo{address}{Berlin}, \bibinfo{year}{2005}).

\bibitem[{\citenamefont{Fabiny et~al.}(1993)\citenamefont{Fabiny, Colet, Roy,
  and Lenstra}}]{Fabiny:Lasers}
\bibinfo{author}{\bibfnamefont{L.}~\bibnamefont{Fabiny}},
  \bibinfo{author}{\bibfnamefont{P.}~\bibnamefont{Colet}},
  \bibinfo{author}{\bibfnamefont{R.}~\bibnamefont{Roy}}, \bibnamefont{and}
  \bibinfo{author}{\bibfnamefont{D.}~\bibnamefont{Lenstra}},
  \bibinfo{journal}{Physical Review} \textbf{\bibinfo{volume}{A47}},
  \bibinfo{pages}{4287} (\bibinfo{year}{1993}).

\bibitem[{\citenamefont{Jr. et~al.}(1997)\citenamefont{Thornburg, M\"oller, Roy,
  Carr, Li, and Erneux}}]{Roy:Lasers}
\bibinfo{author}{\bibfnamefont{K.~S.}~\bibnamefont{Thornburg}},
  \bibinfo{author}{\bibfnamefont{M.}~\bibnamefont{M\"oller}},
  \bibinfo{author}{\bibfnamefont{R.}~\bibnamefont{Roy}},
  \bibinfo{author}{\bibfnamefont{T.}~\bibnamefont{Carr}},
  \bibinfo{author}{\bibfnamefont{R.-D.} \bibnamefont{Li}}, \bibnamefont{and}
  \bibinfo{author}{\bibfnamefont{T.}~\bibnamefont{Erneux}},
  \bibinfo{journal}{Physical Review} \textbf{\bibinfo{volume}{E55}},
  \bibinfo{pages}{3865} (\bibinfo{year}{1997}).

\bibitem[{\citenamefont{Ciofini et~al.}(1993)\citenamefont{Ciofini, Politi, and
  Meucci}}]{Politi:Nonlinearities}
\bibinfo{author}{\bibfnamefont{M.}~\bibnamefont{Ciofini}},
  \bibinfo{author}{\bibfnamefont{A.}~\bibnamefont{Politi}}, \bibnamefont{and}
  \bibinfo{author}{\bibfnamefont{R.}~\bibnamefont{Meucci}},
  \bibinfo{journal}{Physical Review A} \textbf{\bibinfo{volume}{48}},
  \bibinfo{pages}{605} (\bibinfo{year}{1993}).

\bibitem[{\citenamefont{DeShazer et~al.}(2003)\citenamefont{DeShazer,
  Garcia-Ojalvo, and Roy}}]{DeShazer:Lasers}
\bibinfo{author}{\bibfnamefont{D.~J.} \bibnamefont{DeShazer}},
  \bibinfo{author}{\bibfnamefont{J.}~\bibnamefont{Garcia-Ojalvo}},
  \bibnamefont{and} \bibinfo{author}{\bibfnamefont{R.}~\bibnamefont{Roy}},
  \bibinfo{journal}{Physical Review E} \textbf{\bibinfo{volume}{67}},
  \bibinfo{pages}{036602} (\bibinfo{year}{2003}).

\bibitem[{\citenamefont{Glendinning}(1994)}]{Glendinning:StabilityInstabilityC%
haos}
\bibinfo{author}{\bibfnamefont{P.}~\bibnamefont{Glendinning}},
  \emph{\bibinfo{title}{Stability, Instability and Chaos: an introduction to
  the theory of nonlinear differential equations}}
  (\bibinfo{publisher}{Cambridge University Press},
  \bibinfo{address}{Cambridge}, \bibinfo{year}{1994}).

\bibitem[{\citenamefont{Steele and Henderson}(1992)}]{Steele:Plankton}
\bibinfo{author}{\bibfnamefont{J.~H.} \bibnamefont{Steele}} \bibnamefont{and}
  \bibinfo{author}{\bibfnamefont{E.~W.} \bibnamefont{Henderson}},
  \bibinfo{journal}{J.~Plankton Research} \textbf{\bibinfo{volume}{14}},
  \bibinfo{pages}{157} (\bibinfo{year}{1992}).

\bibitem[{\citenamefont{Ludwig et~al.}(1978)\citenamefont{Ludwig, Jones, and
  Holling}}]{Ludwig:Budworm}
\bibinfo{author}{\bibfnamefont{D.}~\bibnamefont{Ludwig}},
  \bibinfo{author}{\bibfnamefont{D.~D.} \bibnamefont{Jones}}, \bibnamefont{and}
  \bibinfo{author}{\bibfnamefont{C.~S.} \bibnamefont{Holling}},
  \bibinfo{journal}{Journal of animal ecology} \textbf{\bibinfo{volume}{47}},
  \bibinfo{pages}{315} (\bibinfo{year}{1978}).

\bibitem[{\citenamefont{Blasius et~al.}(1999)\citenamefont{Blasius, Huppert,
  and Stone}}]{Blasius:Synchronization}
\bibinfo{author}{\bibfnamefont{B.}~\bibnamefont{Blasius}},
  \bibinfo{author}{\bibfnamefont{A.}~\bibnamefont{Huppert}}, \bibnamefont{and}
  \bibinfo{author}{\bibfnamefont{L.}~\bibnamefont{Stone}},
  \bibinfo{journal}{Nature} \textbf{\bibinfo{volume}{399}},
  \bibinfo{pages}{354} (\bibinfo{year}{1999}).

\bibitem[{\citenamefont{Gentleman et~al.}(2003)\citenamefont{Gentleman,
  Leising, Frost, Storm, and Murray}}]{Gentleman:Responses}
\bibinfo{author}{\bibfnamefont{W.}~\bibnamefont{Gentleman}},
  \bibinfo{author}{\bibfnamefont{A.}~\bibnamefont{Leising}},
  \bibinfo{author}{\bibfnamefont{B.}~\bibnamefont{Frost}},
  \bibinfo{author}{\bibfnamefont{S.}~\bibnamefont{Storm}}, \bibnamefont{and}
  \bibinfo{author}{\bibfnamefont{J.}~\bibnamefont{Murray}},
  \bibinfo{journal}{Deep Sea Research {II}} \textbf{\bibinfo{volume}{50}},
  \bibinfo{pages}{2847} (\bibinfo{year}{2003}).

\bibitem[{\citenamefont{Edwards and Bees}(2001)}]{Edwards:Exponent}
\bibinfo{author}{\bibfnamefont{A.~M.} \bibnamefont{Edwards}} \bibnamefont{and}
  \bibinfo{author}{\bibfnamefont{M.~A.} \bibnamefont{Bees}},
  \bibinfo{journal}{Chaos, Solitons and Fractals}
  \textbf{\bibinfo{volume}{12}}, \bibinfo{pages}{289} (\bibinfo{year}{2001}).

\bibitem[{\citenamefont{Abrams and Ginzburg}(2000)}]{Abrams:Debate}
\bibinfo{author}{\bibfnamefont{P.~A.} \bibnamefont{Abrams}} \bibnamefont{and}
  \bibinfo{author}{\bibfnamefont{L.~R.} \bibnamefont{Ginzburg}},
  \bibinfo{journal}{Trends in Ecology and Evolution}
  \textbf{\bibinfo{volume}{15}}, \bibinfo{pages}{337} (\bibinfo{year}{2000}).

\bibitem[{\citenamefont{May}(1973)}]{May:ComplexityStability}
\bibinfo{author}{\bibfnamefont{R.~M.} \bibnamefont{May}},
  \emph{\bibinfo{title}{Stability and Complexity in model ecosystems}}
  (\bibinfo{publisher}{Princeton University Press},
  \bibinfo{address}{Princeton}, \bibinfo{year}{1973}).

\bibitem[{\citenamefont{Haydon}(2000)}]{Haydon:Stability}
\bibinfo{author}{\bibfnamefont{D.}~\bibnamefont{Haydon}},
  \bibinfo{journal}{Ecology} \textbf{\bibinfo{volume}{81}},
  \bibinfo{pages}{2631} (\bibinfo{year}{2000}).

\bibitem[{\citenamefont{Baurmann et~al.}()\citenamefont{Baurmann, Gross, and
  Feudel}}]{spatialpaper}
\bibinfo{author}{\bibfnamefont{M.}~\bibnamefont{Baurmann}},
  \bibinfo{author}{\bibfnamefont{T.}~\bibnamefont{Gross}}, \bibnamefont{and}
  \bibinfo{author}{\bibfnamefont{U.}~\bibnamefont{Feudel}},
  \bibinfo{note}{submitted to Journal of Theoretical Biology}.

\end{thebibliography}
\end{document}